\documentclass[12pt]{iopart}

\usepackage{graphicx}
\usepackage{bm}
\usepackage{color}

\expandafter\let\csname equation*\endcsname\relax 
\expandafter\let\csname endequation*\endcsname\relax 
\usepackage{amsmath}

\usepackage{iopams}
\usepackage{subfigure}

\newcommand{\pder}[2]{\frac{\partial{#1}}{\partial{#2}}}

\newcommand\sig{\sigma({\bf x},t)}

\begin{document}
 
\title{Mass fluctuations and diffusion in time-dependent random environments}

\author{Giorgio Krstulovic, Rehab Bitane, and J{\'e}r{\'e}mie Bec} 

\address{Laboratoire Lagrange UMR7293, Universit\'e de Nice Sophia-Antipolis, CNRS, Observatoire de la C\^ote d'Azur, B.P. 4229, 06304 Nice Cedex 4, France}

\begin{abstract}
A mass ejection model in a time-dependent random environment with both temporal and spatial correlations is introduced. When the environment has a finite correlation length, individual particle trajectories are found to diffuse at large times with a displacement distribution that approaches a Gaussian. The collective dynamics of diffusing particles reaches a statistically stationary state, which is characterized in terms of a fluctuating mass density field. The probability distribution of density is studied numerically for both smooth and non-smooth scale-invariant random environments. A competition between trapping in the regions where the ejection rate of the environment vanishes and mixing due to its temporal dependence leads to large fluctuations of mass. These mechanisms are found to result in the presence of intermediate power-law tails in the probability distribution of the mass density. For spatially differentiable environments, the exponent of the right tail is shown to be universal and equal to $-3/2$. However,  at small values, it is found to depend on the environment. Finally, spatial scaling properties of the mass distribution are investigated. The distribution of the coarse-grained density is shown to posses some rescaling properties that depend on the scale, the amplitude of the ejection rate, and the H\"older exponent of the environment.
\end{abstract}
\maketitle

\section{Introduction}

Many different situations found in nature take place in an environment where spatial fluctuations occur at so small length scales that they contribute only through an averaged effect on macroscopic processes. The environment is then usually modeled by introducing some disorder, like in the Sherrington and Kirkpatrick spin-glass model for magnetization~\cite{Sherrington1975}. In addition the time variations of the environment are often much slower than those of the mechanism of interest (such as diffusion, transport or wave propagation). This leads on to consider a quenched disorder, as in the case of directed polymers \cite{HalpinHealy1995215,ImbrieandSpencer_1988} and of wave propagation in random media \cite{Howe_1971}.  As a consequence, the discrete version of diffusive processes in quenched random media, known as random walks in random environment (RWRE) \cite{Hughes_RWRE}, have been largely studied by mathematicians and physicists. One usually considers a lattice with fixed random transition probabilities between sites and study  the behavior of a random walk on it.  One of the main questions of interest has been to determine under which assumptions such RWRE are transient or recurrent (that is whether they escape to infinity or indefinitely come back to their starting position). For a time-independent random environment with no space correlations, it has been shown that randomizing the environment slowdowns the diffusive properties of the random walk \cite{solomon1975random}. Furthermore, under some precise assumptions, Ya.~Sinai \cite{sinai1982limit} proved that one-dimensional symmetric nearest-neighbor random walks in an uncorrelated environment are sub-diffusive and scale as $X_t\sim \log^2{t}$ when $t\to\infty$. B.~Derrida and Y.~Pomeau~\cite{derrida1982classical} demonstrated that when the assumption of symmetry is relaxed, the random walk escapes as $X_t\sim t^\alpha$, with $0<\alpha<1$. J.~Bricmont and A.~Kupiainen~\cite{bricmont1991random} showed that the upper critical dimension of random walks in a time-independent random environment is $2$, so that for $d>2$ RWRE are diffusive. More recently, a large effort has been devoted to prove central-limit theorems or large-deviations principles for quenched and annealed disorders \cite{sznitman2002topics,zeitouni2004part,Varadhan2004}. Such work generally  assume that the environment has no (or very short) spatial correlations and that it is time-independent. Only little is known for the case of time-dependent environments.  In such settings and under some rather general assumptions, it was shown that random walks are always diffusive \cite{boldrighini2004random} and that central limit theorems generally hold \cite{berad2004almost}. 

We are here interested in situations where the timescales and length scales of diffusion are comparable or longer than those at which the environment fluctuates.  A clear instance where this occurs is turbulent transport whose modeling is a key issue in industrial and environmental sciences. The classical models used for instance in engineering and meteorology are based on an ``eddy-diffusivity'' approach (see, e.g., \cite{Frisch:TurbuBook}), which in general requires a clear scale separation between the turbulence and large-scale variations of the mean velocity.  Homogenization techniques can then be used to show that the averaged concentration field follows an effective advection-diffusion equation with an advecting velocity and a diffusion coefficient tensor that depend on the slow variables~\cite{goudon2004homogenization}. The large-scale mixing originating from small-scale fluctuations acts only through the diffusive term, which, by the maximum principle, cannot be responsible for the creation of large concentrations. The concentrations observed in compressible flows can thus only come from the effective advection term. As we will see in this paper, the situation can be rather different when interested in other types of compressible transport.

Let us consider for instance species whose dynamics is dissipative in the full position-velocity phase space. We assume that the trajectory of a particle obeys the Newton equation
\begin{eqnarray}
\dot{X_t}=V_t,\hspace{5mm}\dot{V}_t=-\mu V_t+\frac{1}{\sqrt{\varepsilon}}f(X_t,t,{t}/{\varepsilon}).\label{Eq:motiv}
\end{eqnarray}
The forces acting on the particle are a viscous drag and an external force $f$ that depends on both space and two different time scales. We have dropped here the vectorial notations but all the following considerations can be easily generalized to any dimension. Such an equation describes for instance the dynamics of a heavy inertial particle in a velocity field that varies over two time scales. Also, one could consider a large particle embedded in a time and space-dependent thermal bath. The fast time scale $\varepsilon$ can be interpreted as the typical time of momentum exchange between the particle and its environment. As in the Einstein original work about Brownian motion, the time-scale $\varepsilon$ is much smaller than the viscous damping time $1/\mu$. In the limit $\varepsilon \mu \to 0$, the force $f$ can be approximated by a Gaussian noise with correlation
\begin{equation}
  C(t,t') = (1/\varepsilon) \langle f(X_t,t,t/\varepsilon)f(X_{t'},t',t'/\varepsilon)\rangle_\varepsilon \simeq 2\Gamma\mu^2\sigma^2(X_t,t)\delta(t-t'),
\end{equation}
where the average $\langle\cdot\rangle_\varepsilon$ is with respect to the fast time variable. The spatio-temporal variations of $\sigma^2$ can be interpreted as a non-homogenous temperature field in the thermal bath. We next consider the limit when the response time $1/\mu$ is much shorter than the slow time scale. This introduces a new fast time scale $\mathrm{O}(1/\mu)$ over which the particle velocity fluctuates. We see that
\begin{eqnarray}
  \langle V_t\,V_{t'}\rangle_{\varepsilon}&=&\int\limits_{-\infty}^t \int\limits_{-\infty}^{t'}  C(s,s') \mathrm{e}^{\mu(s+s'-t-t')} \mathrm{d}s\,\mathrm{d}s' \nonumber \\
  &=& \int\limits_{-\infty}^{t\land t'}2\Gamma\mu^2\sigma^2(X_s,s)\mathrm{e}^{\mu(2s-t-t')} \mathrm{d}s \approx \Gamma\mu\, \sigma^2(X_{t\land t'},t\land t')      \mathrm{e}^{-\mu|t-t'|}
\label{Eq:correlationLangevin}
\end{eqnarray}
where the last equality is obtained integrating by parts and neglecting exponentially small terms. Then, taking the limit $\mu\to\infty$ in \eqref{Eq:correlationLangevin} we obtain $\langle V_t\,V_{t'}\rangle= 2\Gamma\sigma^2(X_t,t) \delta(t-t')$. The particle position thus satisfies the stochastic equation
\begin{equation}
\mathrm{d}X_t=\sqrt{2\Gamma}\,\sigma(X_t, t)\,\mathrm{d}W_t,
\end{equation}
where $W_t$ is the Wiener process and where the product has to be interpreted in the It\^o sense. The averaged particle concentration $\rho$ is then a solution to the associated forward Kolmogorov equation, namely
\begin{equation}
\frac{\partial\rho}{\partial t} = \Gamma \frac{\partial^2}{\partial x^2} \left[ \sigma^2(x, t)\,\rho\right].
\end{equation}
The solutions to such an equation have a very different behavior than those to the standard diffusion equation where on the right-hand side $\sigma^2$ would appear in between the two spatial derivatives. It is for instance clear that the mass is going to accumulate at the zeros of $\sigma$.  Indeed, suppose $\sigma(x,t) \simeq C x$ in the vicinity of $x=0$. At leading order, the flux reads $J = -\Gamma (\partial/\partial x) ( \sigma^2\rho) \simeq -2\Gamma C^2 x \rho$, which is positive for $x<0$ and negative for $x>0$, leading to a permanent mass flux toward $x=0$. As we will see later, the zeros, their densities and their lifetimes play a crucial role in the statistical properties of the density field. 

The paper is organized as follows: in section \ref{Sec:DescripModel} we introduce a time and space continuous model of diffusion in a random environment based on an ejection model in a discrete $D$-dimensional lattice. The model is reinterpreted in terms of particle dynamics obeying an It\^o differential equation with a multiplicative noise and no drift. The random environment is then explicitly defined and its general spatial and temporal properties are discussed. Section \ref{Sec:DiffProp} is devoted to the study of individual trajectories of the It\^o diffusion in the case of smooth and non-smooth environments. When the environment has a finite correlation length, particles are found to have standard diffusion properties at large times with a displacement distribution that approaches a Gaussian. In section \ref{Sec:DensityFluct} we define the density of mass and study the fluctuations of the density in smooth and non-smooth random environments. The locations where the environment ejection rate vanishes are shown to play a crucial role on the large fluctuations of density. Spatial scaling properties of the mass distribution are then investigated in \ref{Sec:ScaleInv}. We show that the distribution of the coarse-grained density possesses some rescaling properties that depend on the scale, the amplitude of the ejection rate, and the H\"older exponent of the environment. Finally, Section \ref{Sec:Concl} presents some concluding remarks.

\section{Description of the model\label{Sec:DescripModel}}

Let us consider the $D$-dimensional periodical lattice of fundamental size $\Delta x$ and period $L=N \Delta x$. This lattice defines a tiling on which we consider the following discrete-time dynamics. The cell indexed by ${\bf i}$ contains at time $n\Delta t$ a mass $\rho_{\bf i}(n)$. Then, between times $n\Delta t$ and $(n+1)\Delta t$, a fraction $0<\gamma_{{\bf i}}(n)<1/(2D)$ of this mass is ejected from the cell ${\bf i}$ to one of its $2D$ neighbors. For a given set of ejection rates $\{\gamma_{\bf i}(n)\}_{{\bf i}\in\{1,\ldots,N\}^D}$  the variables $\{\rho_{\bf i}(n)\}_{{\bf i}\in\{1,\ldots,N\}^D}$ define a Markov chain whose master equation is given by
\begin{equation}
\rho_{\bf i}(n+1)=\left[1-2D\,\gamma_{{\bf i}}(n)\right]\rho_{\bf i}(n)+\sum_{{\bf j}\in{\rm \mathcal{N}_{\bf i} }}\gamma_{{\bf j}}(n)\rho_{{\bf j}}(n),\label{Eq:Model_Discrete}
\end{equation}
where $\mathcal{N}_{\bf i} $ are the neighbor cells of ${\bf i}$.
It is clear that, because of periodic boundary conditions, the total mass $M=\sum_{\bf i} \rho_{\bf i}$ is conserved.
Equation \eqref{Eq:Model_Discrete} describes a mass-ejection process in a time-dependent non-homogeneous environment determined by the ejection rates $\gamma_{{\bf i}}(n)$. A similar model has been studied in \cite{Bec:NJP2007} in the case when the $\gamma_{{\bf i}}(n)$ take two values, $\gamma$ and $0$ and are uncorrelated in both space and time. This model was motivated by the study of inertial particles clustering in turbulent flows and was expected to mimic heavy particle ejection from coherent vortical structures by centrifugal forces. 

Turning back to the ejection model (\ref{Eq:Model_Discrete}), we now consider the continuous limit $\Delta x\to 0$ and $\Delta t\to 0$. To take this limit, we suppose that the ratio $\Delta x^2/\Delta t$ is kept fixed.  We denote by $\Gamma\sigma^2({\bf x},t)$ the continuous limit of $\gamma_{{\bf x}/\Delta x}(t/\Delta t)$, where the coefficient $\Gamma$ is the typical amplitude of the ejection rate and $\sigma$ is an order-one time and space continuous function. In this limit, equation \eqref{Eq:Model_Discrete} becomes
\begin{equation}
\pder{\rho({\bf x},t)}{t}=\Gamma\,\nabla^2\left[\sigma^2({\bf x},t)\rho({\bf x},t)\right].\label{Eq:ModelD}
\end{equation}
As mentioned in the introduction, in contrast with a standard diffusion equation coming from Fick's first law, the differential operator $\nabla^2[\sigma^2({\bf x},t)\,\cdot ]$ in the right-hand side of equation \eqref{Eq:ModelD} is not positive-definite. There is no maximum principle and the solution is expected to behave very differently from those to standard diffusion equations.

Equation (\ref{Eq:ModelD}) is the Fokker-Planck (forward Kolmogorov) equation associated to the It\^o stochastic differential equation
\begin{equation}
  \mathrm{d} {\bf X}_t = \sqrt{2\Gamma}\,{\sigma}({\bf X}_t,t)\,\mathrm{d}{\bf W}_t,
  \label{eq:sde}
\end{equation}
where ${\bf W}_t$ is the $D$-dimensional Wiener process. For $t>s$, we define the (forward) transition probability density
\begin{equation}
  p({\bf x},t\,|\,{\bf y},s) = \left\langle \delta({\bf X_t}-{\bf x}) \,|\, {\bf X_s=y}\right\rangle,
\end{equation}
where $\langle\cdot\rangle$ designates averages with respect to the realizations of ${\bf W}_t$. Then the solutions to (\ref{Eq:ModelD}) trivially satisfy for all $t>s$
\begin{equation}
  \rho({\bf x},t) = \int p({\bf x},t\,|\,{\bf y},s)\,\rho({\bf y},s) \mathrm{d}{\bf y}.
\end{equation}
The model described by the diffusion equation \eqref{Eq:ModelD} can thus be reinterpreted in terms of a random walk in a random environment (RWRE). Indeed the discrete version of the process given by (\ref{eq:sde}) corresponds to a random walk on the $D$-dimensional lattice, with a probability to hop from the site $\bf i$ to one of its neighbors equal to  $2D\,\gamma_{\bf i}(n)$.

Let us now specify the statistical properties of the random environment. We assume that $L=2\pi$ and that $\sigma({\bf x},t)$ is $2\pi$-periodic in space. The environment is then entirely determined in terms of its Fourier series that we write as
\begin{equation}
  \sigma({\bf x},t) = \sum_{|{\bf k}|}a_{\bf k}\,\chi_{\bf k}(t/\tau_{\bf k})
  \,\mathrm{e}^{i{\bf k}\cdot{\bf x}},\label{Eq:defSigma}
\end{equation}
where the $a_{\bf k}$'s are positive real amplitudes and the $\tau_{\bf k}$'s are scale-dependent characteristic times. The Fourier modes $\chi_{\bf k}$ satisfy $\chi_{-{\bf k}}(t) = \overline{\chi}_{\bf k}(t)$ and $\chi_{\bf 0}\equiv 0$. They are independent Gaussian processes with unit variance and unit correlation time. We choose them as complex Ornstein--Uhlenbeck processes that solve the stochastic differential equation \cite{OrnsteinUhlenbeck}
\begin{equation}
  \mathrm{d}\chi_{\bf k}(s) = - \chi_{\bf k}\mathrm{d}s +
  \sqrt{2}\,\mathrm{d}B_{\bf k}(s),\label{Eq:AmpSigma}
\end{equation}
where the $B_{\bf k}$'s are independent one-dimensional complex Wiener processes.

We now prescribe scale-invariance properties for the random environment. We want $|\sigma({\bf x}+{\bf r}) - \sigma({\bf x})| \sim |{\bf r}|^h$, where $h>0$ controls the regularity of the ejection rate. When $h<1$, it is the spatial H\"older exponent of the ejection rate. This amounts to assuming that the Fourier mode amplitudes behave as $a_k =\frac{1}{2} k^{-1/2-h}$. Temporal scale invariance is set by assuming $\tau_k = k^{-\beta}$ with $\beta>0$. The exponent $\beta$ relates time dependence to spatial scale invariance; the case $\beta=0$ corresponds to ${\sigma}$ having a unique correlation time and $\beta\to\infty$ to white-noise. As we shall see below, when $h<1$, dimensional analysis motivates the choice $\beta = 2-2h$.

The numerical results presented in this work are obtained by two different manners. The density evolution is obtained by solving equation \eqref{Eq:ModelD} with a second-order finite-difference discretization of the Laplacian and a semi-implicit Euler temporal scheme. This numerical scheme ensures the conservation of the total mass with a high accuracy. In this work numerical resolutions vary from $N=512$ to $N=8192$ collocation points and the time-step is chosen small enough to well resolve the fastest time scale of the problem.  For particle trajectories, the stochastic equations  \eqref{eq:sde} and \eqref{Eq:AmpSigma} are solved by a standard stochastic Euler scheme and ensemble averages are obtained by Monte-Carlo methods. In both cases, we expect the error made on the solution to act as a numerical diffusion with a constant proportional to the time step.

\begin{figure}[h!]
\centering
\subfigure[$D=1$]{\includegraphics[height=6cm]{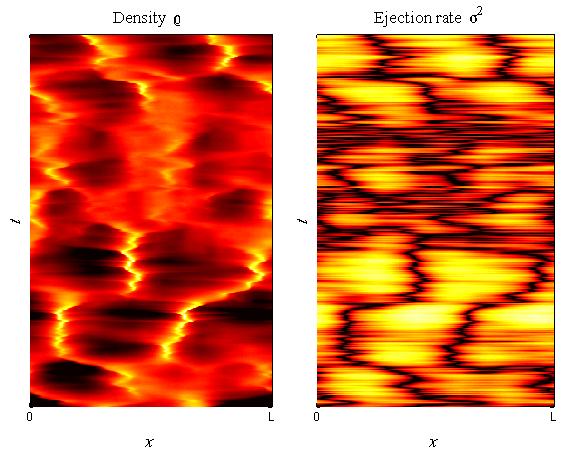} \label{fig:snapshot_a}}
\qquad
\subfigure[$D=2$]{\includegraphics[height=6cm]{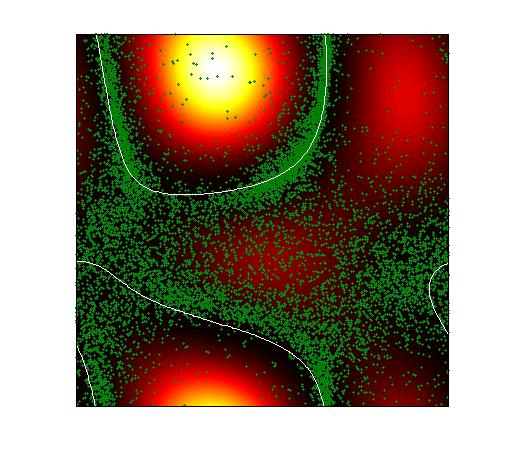} \label{fig:snapshot_b}}
\caption{\label{fig:snapshot} (a) Left: space-time plot of $\rho(x,t)$  (high densities in yellow and low in black) for $h=1$ in dimension $D=1$ a Right: corresponding evolution of the ejection rate $\sigma^2(x,t)$ during the same realization; low values are in dark and high values in yellow.  (b) Snapshot of the position of particles obeying \eqref{eq:sde} (in green) for $D=2$, together with $\sigma^2(x,t)$. Low values of $\sigma^2(x,t)$ are in dark and high values in yellow; contour lines of $\sigma=0$ are shown in white.}
\end{figure}
Figure \ref{fig:snapshot_a} show the temporal evolution of the density and of the ejection rate  $\sigma^2(x,t)$, respectively, for a typical realization in dimension $D=1$. Figure \ref{fig:snapshot_b} shows a snapshot of the random environment $\sigma$ and of the position of a set of particles obeying equation \eqref{eq:sde} in $D=2$. It is apparent on figure \ref{fig:snapshot} that the high density zones are located in the vicinity of the zeros of $\sig$.  From the right of figure \ref{fig:snapshot_a} we observe that the zeros of $\sigma$  follow random paths. Their role on the statistical properties of the density of mass and on the diffusion properties of particles will be further discussed trough this work.

The distribution and time-evolution of these zeros strongly depend on the parameter $h$ characterizing the environment. Let us first consider the case of large $h$. On figure \ref{fig:snapshot_a} it is apparent that the zeros typically appear by pairs and then diffuse and separate until they merge together or with another zero. Without loss of generality, we can assume that the ejection rate has only two modes and takes the simple form
\begin{equation}
  \sigma(x,t) = \sigma_r(t)\,\cos x - \sigma_i(t)\,\sin x,\label{Eq:Sigma1}
\end{equation}
with $\sigma_r+\mathrm{i}\,\sigma_i = 2\chi_1$. Using the It\^o formula it is possible to show that $\sigma$ can be equivalently rewritten as $\sigma(x,t) = A_t\,\sin\left( x - \Phi_t\right)$, where the stochastic processes $A_t$ and $\Phi_t$ satisfy
\begin{eqnarray}
    \mathrm{d}A_t &=& -\left(A_t-\frac{1}{A_t}\right)\mathrm{d}t + \sqrt{2}\,\mathrm{d}B_t^A\hspace{.25cm}, \hspace{.25cm} \mathrm{d}\Phi_t=\frac{\sqrt{2}}{A_t}\mathrm{d}B_t^\Phi ,\label{Eq:Ampzeros}
\end{eqnarray}
where $B_t^A$ and $B_t^\Phi$  are uncorrelated Wiener processes. It is clear that the amplitude $A_t$ fluctuates around $A=1$ and that its correlation time is order one. It follows from equation \eqref{Eq:Ampzeros} that, for typical values of $A_t$, the phase $\Phi_t$, and thus the position of the zeros of $\sigma$,  diffuse on a order-one timescale.

When $h<1$, the random environment presents scaling properties  and a very different behavior is expected. We have by construction that the second-order structure function of $\sigma$ behaves as
\begin{equation}
\delta\sigma^2_\ell=\overline{|\sigma(x,t)-\sigma(x+\ell,t)|^2}\sim (\ell/L)^{2h},\label{Eq:def_h}
\end{equation}
where the over-line stands for the ensemble average with respect to the fluctuations of the environment (i.e.\ with respect to the Ornstein--Uhlenbeck processes $\chi_{\bf k}$). In this work, we make the choice of relating space and time correlations by a dimensional argument. The characteristic time $\tau_k$ introduced in equation \eqref{Eq:defSigma}, behaves as a power law, so that the correlation time of $\sigma$ at scale $\ell$ is $\tau_\mathrm{C}(\ell) = (\ell/L)^\beta$. According to equation (\ref{Eq:def_h}) the diffusion timescale associated to the spatial scale $\ell$ behaves as $\tau_\mathrm{D}(\ell) = \ell^2/(\Gamma\delta\sigma^2_\ell)\sim\Gamma^{-1}L^{2h}\ell^{2-2h}$. The ratio $\mathrm{Ku}(\ell) =\tau_\mathrm{C}(\ell)/ \tau_\mathrm{D}(\ell) \simeq \Gamma L^{-(2h+\beta)} \ell^{\beta+2h-2}$ defines a dimensionless space-dependent parameter that is usually called the Kubo number. When $\mathrm{Ku}(\ell)\gg1$, the environment is as if frozen. When $\mathrm{Ku}(\ell)\ll1$, it fluctuates in an almost time-uncorrelated manner. When $\beta\neq 2-2h$, the Kubo number depends on the scale and this breaks any possible scale invariance of the mass distribution. When $\beta = 2-2h$, we have $\mathrm{Ku}(\ell) = \mathrm{Ku} = \Gamma/L^2$, so that scale invariance is possible. Here, we focus on this latter case and will describe the mass concentration properties in terms of the dimensionless parameter $\mathrm{Ku} \propto \Gamma$. Note that the choice of having a scale-independent Kubo number is common in the framework of turbulence and is expected to be relevant to the problem of diffusion of inertial particles in turbulent flows.

Finally, let us do a couple of remarks on the dependence upon the other parameter of the environment, namely the H\"older exponent $h$. Using Parseval's theorem, it is possible to show that  $\overline{\|\sigma\|_{L^2}^2}= \zeta(1+2h)$ and that $\overline{\|\partial_x\sigma\|_{L^2}^2}= \zeta(-1+2h)$, where $\zeta(s)$ is the Riemann zeta function. We thus have that $\overline{\|\sigma\|_{L^2}}<\infty$ when $h>0$ and that $\overline{\|\partial_x\sigma\|_{L^2}}<\infty$ when $h>1$. Note that, at each time $t$ fixed, $\sigma(x,t)$ is a Gaussian variable with the variance of increments  given by equation \eqref{Eq:def_h}.  Therefore  $\sigma(\cdot,t)$ is a fractional Brownian motion of exponent $h$ \cite{mandelbrot:422}. For $h=1/2$ it corresponds to  the standard Brownian motion. For $h>1/2$ the increments of $\sigma$ are not independent and have a positive covariance; the zeros of $\sigma$ are then finite and isolated. For $h<1/2$ the covariance of increments is negative and therefore we expect the number of zeros to become infinite and to accumulate. This different behavior of the random environment will affect the general properties of the diffusion, as we will see in the next sections.

As the Ornstein--Uhlenbeck processes $\chi_{\bf  k}$ are stationary, the fluctuations of the ejection rate $\sigma^2$ is a stationary and homogeneous random field. Hence we expect that, at sufficiently large times, the density distribution reaches a statistically stationary state and numerical simulations indicate that this is indeed the case. The work reported in this paper mainly concerns with the statistical properties of the density field in this large-time asymptotics. Most of the results presented in this work are in dimension $D=1$.

\section{Diffusive properties\label{Sec:DiffProp}}

We now turn to studying the diffusive properties of the solutions $X(t)$ to \eqref{eq:sde}. One has 
\begin{equation}
X(t) = X(0) + \sqrt{2\Gamma} \int_0^t \sigma(X(s),s) \,\mathrm{d}W_s, \label{EQ:integparticles}
\end{equation}
so that 
\begin{equation}
\left\langle [X(t)-X(0)]^2\right\rangle =  2\Gamma \int_0^t \left\langle \sigma^2(X(s),s) \right\rangle \mathrm{d}s, \label{EQ:mean_square_displ}
\end{equation}
where the angular brackets denote average over trajectories. At times much larger than the correlation time of $\sigma^2$ along particle trajectories, the integral becomes a sum of independent random variables, which obeys the law of large numbers. We thus have 
\begin{eqnarray}
\langle [X(t)-X(0)]^2\rangle\sim2D(\Gamma,h)\,t\, \nonumber\\
\mbox{with } D(\Gamma,h) =\Gamma\overline{\langle \sigma^2 (X(t),t)\rangle} =\frac{\Gamma}{L}\int_0^L \overline{\sigma^2 (x,t)\rho(x,t)}\,\mathrm{d}x,\label{Eq:diffLin}
\end{eqnarray}
where the over-line denotes here the average with respect to the environment. The effective diffusion constant $D(\Gamma,h)$ involves an average of the environment along particle trajectories which is equivalent to an average weighted by the mass density $\rho$. As the trajectories are expected to spend a long time in the vicinity of the zeros of $\sigma$, we expect this Lagrangian average to be less than the full Eulerian spatial average. The displacement fluctuations of order less than $\sqrt{t}$ are expected to be given by a central limit theorem. However, larger fluctuations should obey a large deviations principle with a rate function that might not be purely quadratic as it is non trivially related to the Lagrangian properties of the environment. To study these fluctuations, we perform Monte-Carlo simulations of equation \eqref{eq:sde}.

\begin{figure}[h!]
\centering
\includegraphics[width=0.6\columnwidth]{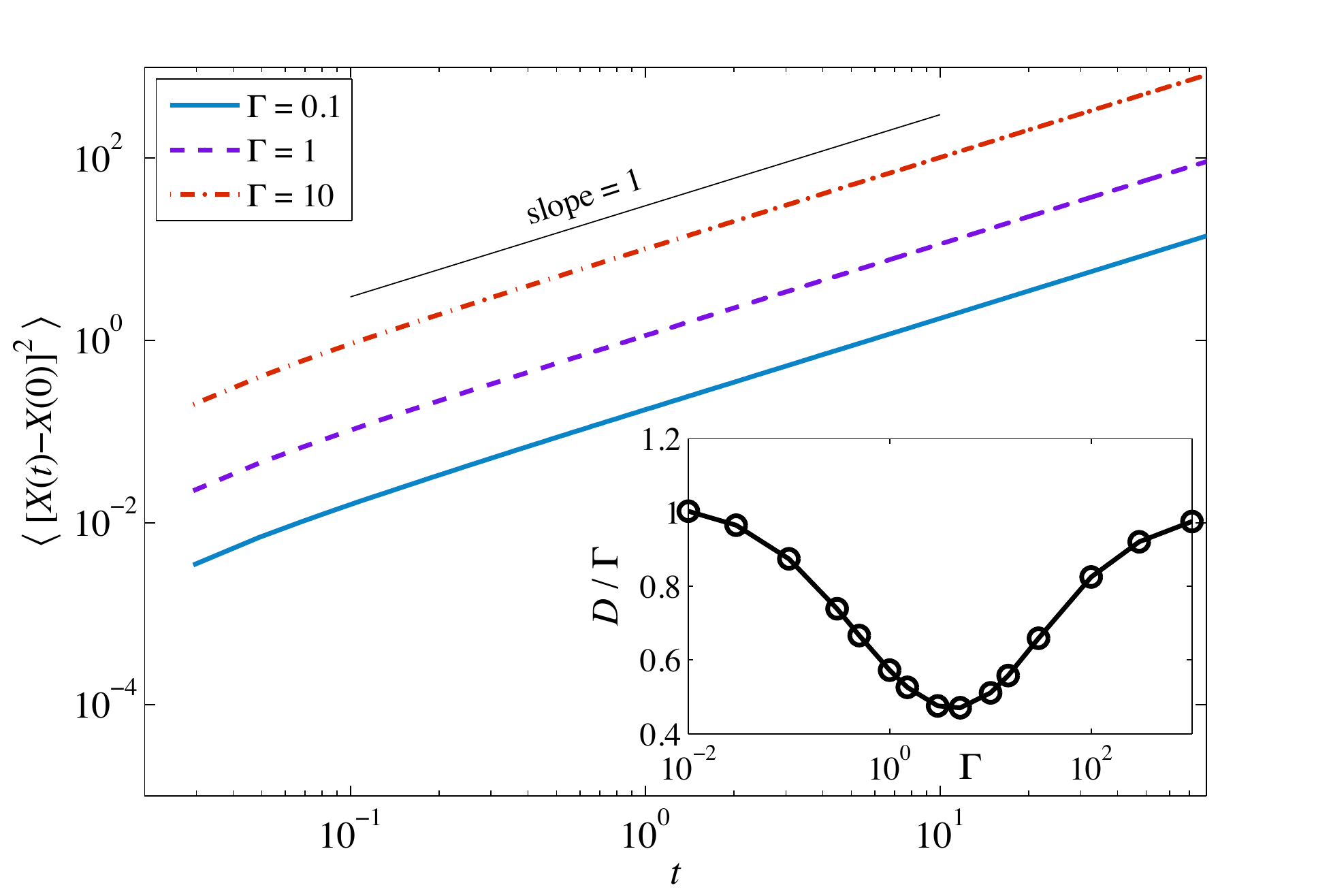}
  \caption{\label{fig:diffuse1} Mean-square displacement  $\overline{\langle (X(t)-X(0))^2\rangle}$ for
    different values of $\Gamma$. Inset: Dependence of the effective diffusion constant \eqref{Eq:diffLin} on $\Gamma$. }
\end{figure}

We first consider the case where the ejection rate $\sigma^2$ is a smooth function of space (i.e.\ $h>1$). With this configuration there are a finite (and small) number of isolated zeros separated by a distance order $L/2=\pi$. A number of simulations of equation \eqref{eq:sde} have been performed for different values of $\Gamma$ using $32$ modes and $h=2$. Figure \ref{fig:diffuse1} shows the mean-square displacement of particles averaged over $1000$ realizations of the diffusion and $20\,000$ realizations of the environment. 
As expected from equation~\eqref{Eq:diffLin}, a linear growth is observed at large times. However the diffusion constant  $D(\Gamma)$ non trivially depends on $\Gamma$, as seen in the inset of figure \ref{fig:diffuse1}. The non-monotonic behavior of $D(\Gamma)/\Gamma$ can be explained by two competing phenomena. In the limit $\Gamma\to 0$ (or equivalently $\mathrm{Ku}\to0$) the environment changes fast enough compared to the characteristic time of diffusion given by $1/\Gamma$. In this limit, we can consider that the particle trajectories sample homogeneously the environment, so that the Lagrangian average can be replaced by the spatial Eulerian average.  This leads to\begin{equation}
 D(\Gamma,h) \simeq \frac{\Gamma}{L}\int \overline{\sigma(x,t)^2} \,\mathrm{d}x = \Gamma\zeta(1+2h) \mbox{ when } \gamma\to 0.
\end{equation}
For $h=2$, we obtain $D(\Gamma,2)/\Gamma \simeq 1.03$, which is in agreement with the value observed on the inset of figure \ref{fig:diffuse1}. As $\Gamma$ increases, trajectories spend a longer time at the zeros of $\sigma$ decreasing the value of $D(\Gamma)/\Gamma$. In the limit of $\Gamma\to\infty$ (i.e.\ $\mathrm{Ku}\to\infty$), the environment can be considered completely frozen and we expect the mass density solving equation \eqref{Eq:ModelD} to be approximatively given  by $\rho(x,t)\sim1/\sigma^2 (x,t)$. From equation (\ref{Eq:diffLin})  we get $D(\Gamma,h)= (\Gamma/L) \int \overline{\sigma^2\rho}\,\mathrm{d}x \approx (\Gamma/L) \int \mathrm{d}x = \Gamma$. Hence $D(\Gamma)/\Gamma=1$ for $\Gamma\to\infty$ as seen on the figure. In this limit the mass is completely concentrated at the zeros of the ejection rate, and the diffusion is carried out by the motions of the laters that follow a dynamics similar to equation \eqref{Eq:Ampzeros}.

\begin{figure}[t!]
\centering
\hspace*{-4pt}
\subfigure[$\Gamma=0.1$]{\includegraphics[width=0.35\columnwidth]{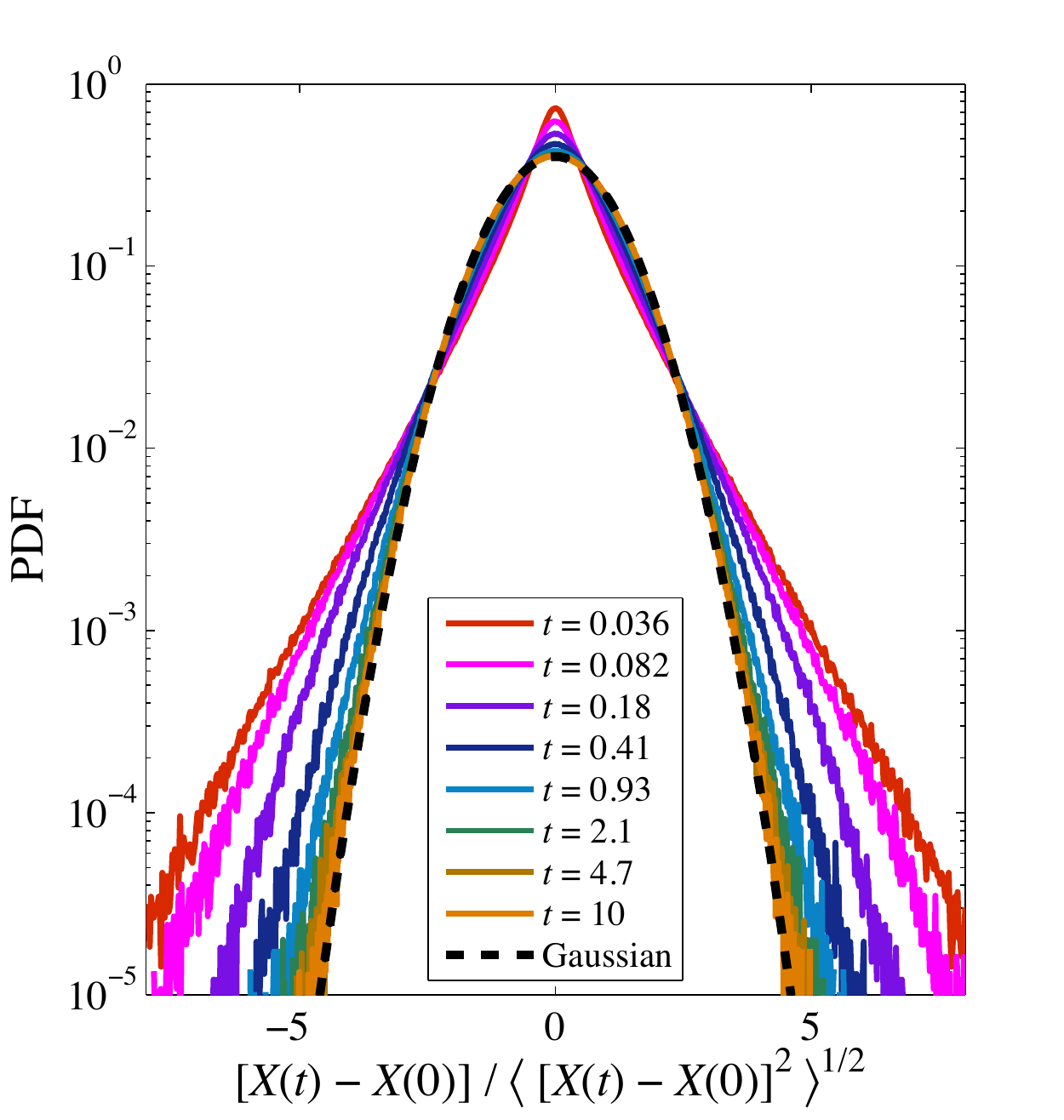}}
\hspace{-18pt}
\subfigure[$\Gamma=1$]{\includegraphics[width=0.35\columnwidth]{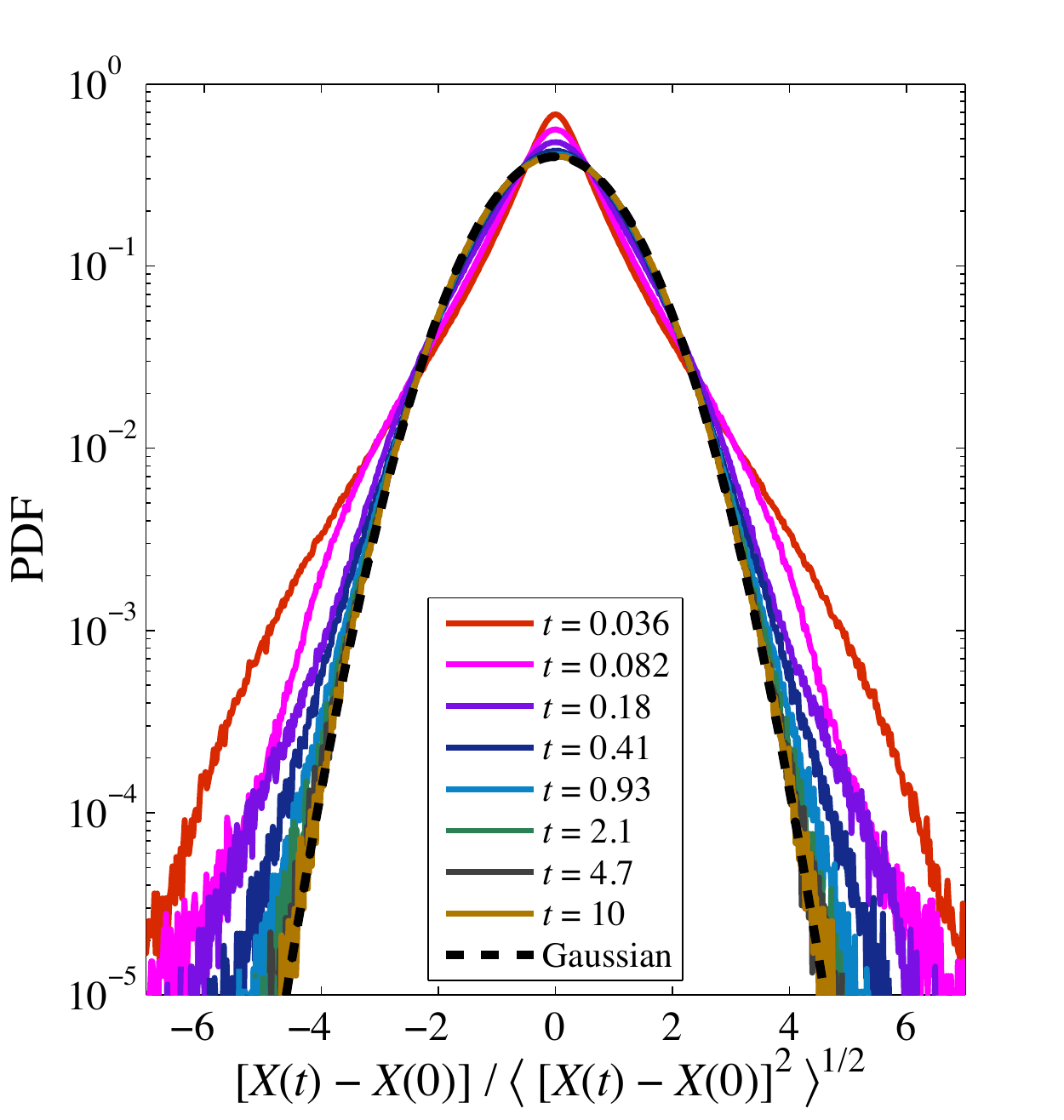}}
\hspace{-18pt}
\subfigure[$\Gamma=10$]{\includegraphics[width=0.35\columnwidth]{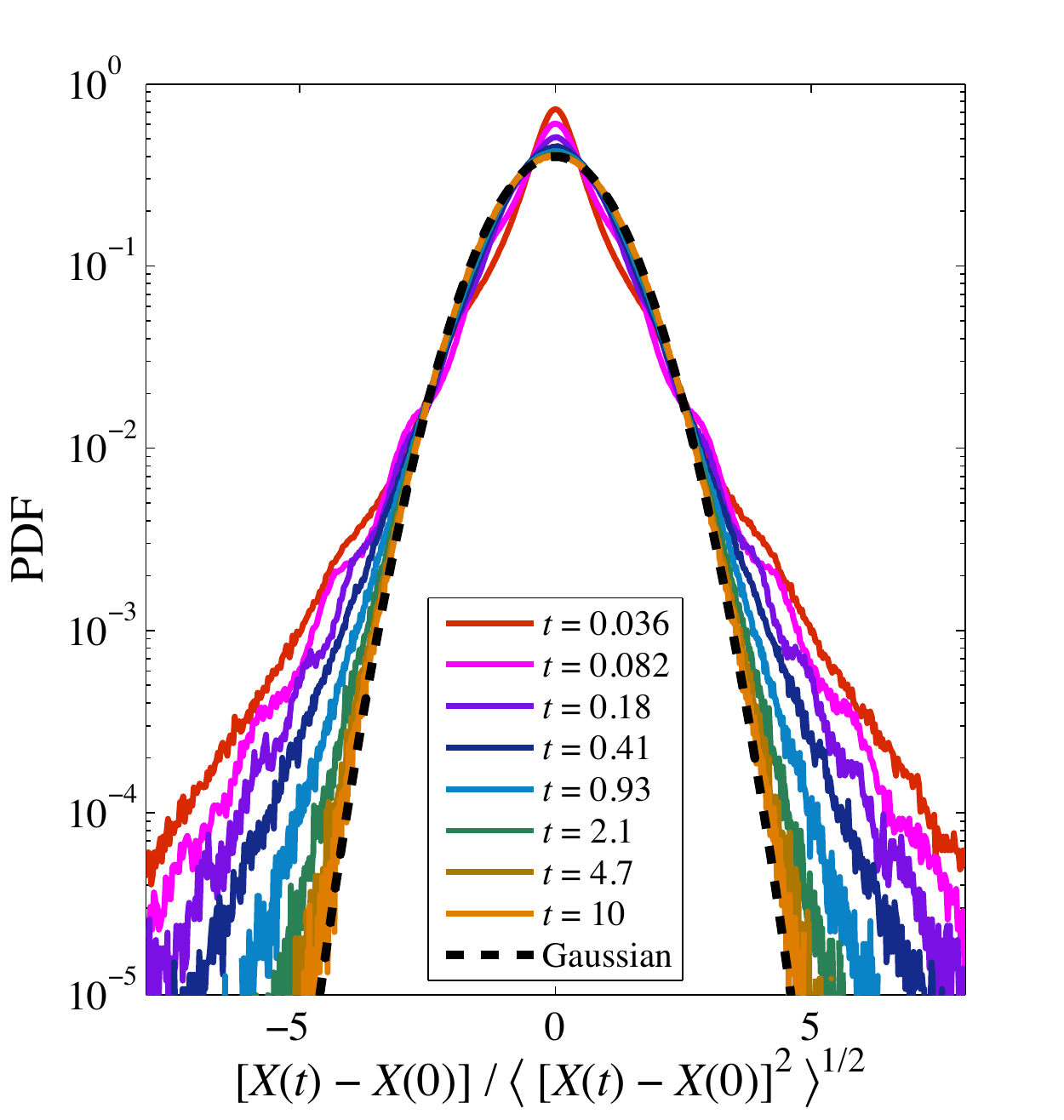}}
\caption{\label{fig:diffuse2} Rescaled PDFs of the displacement $X(t)-X(0)$ for $h=2$, for three different values of $\Gamma$ and for various times, as labeled. The initial condition was chosen in the statistically stationary regime.}
\end{figure}

\begin{figure}[h!]
\centering
\hspace*{-4pt}
\subfigure[]{\includegraphics[width=0.35\columnwidth]{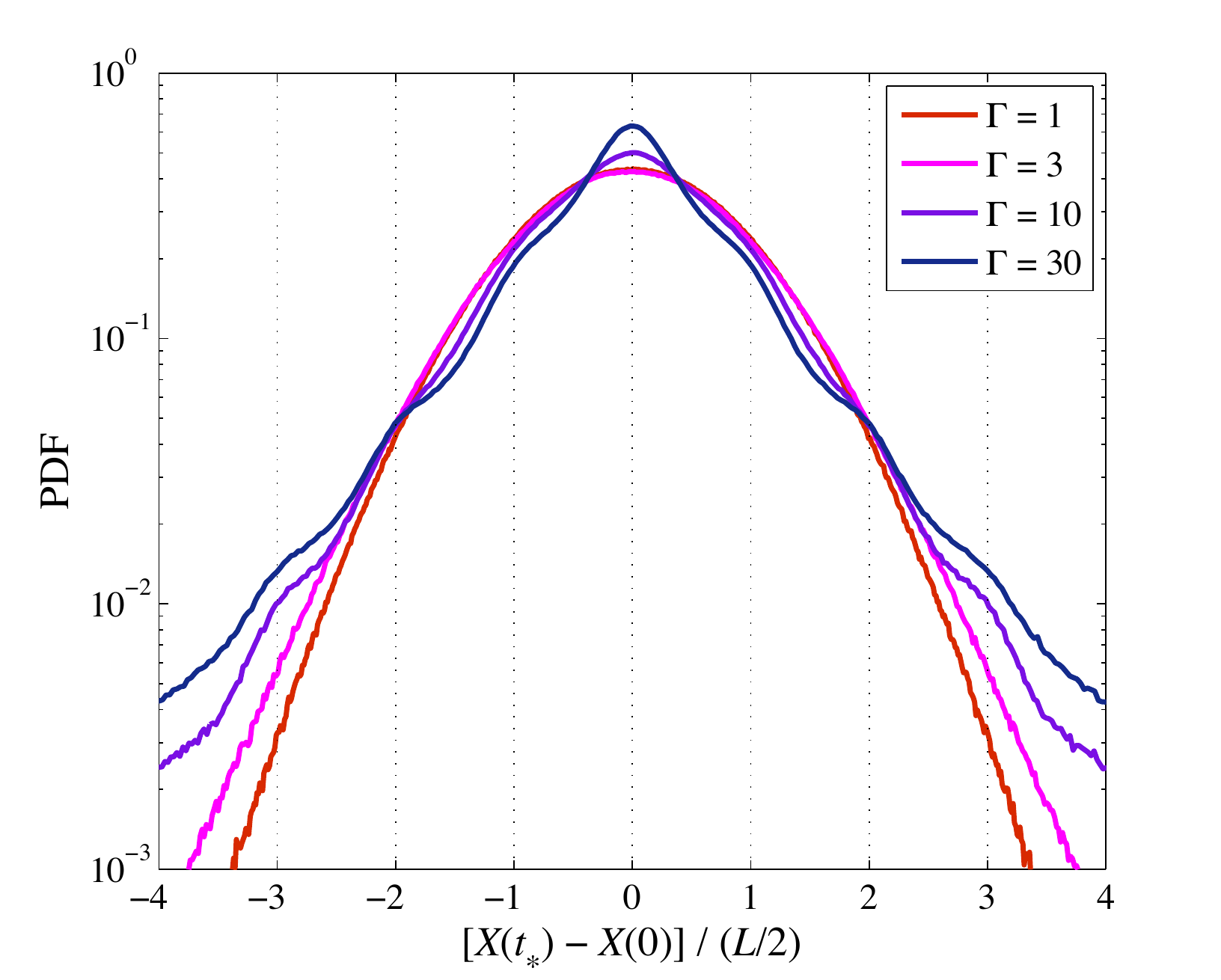}\label{fig:diffuse3_a}}
\hspace{-18pt}
\subfigure[]{\includegraphics[width=0.35\columnwidth]{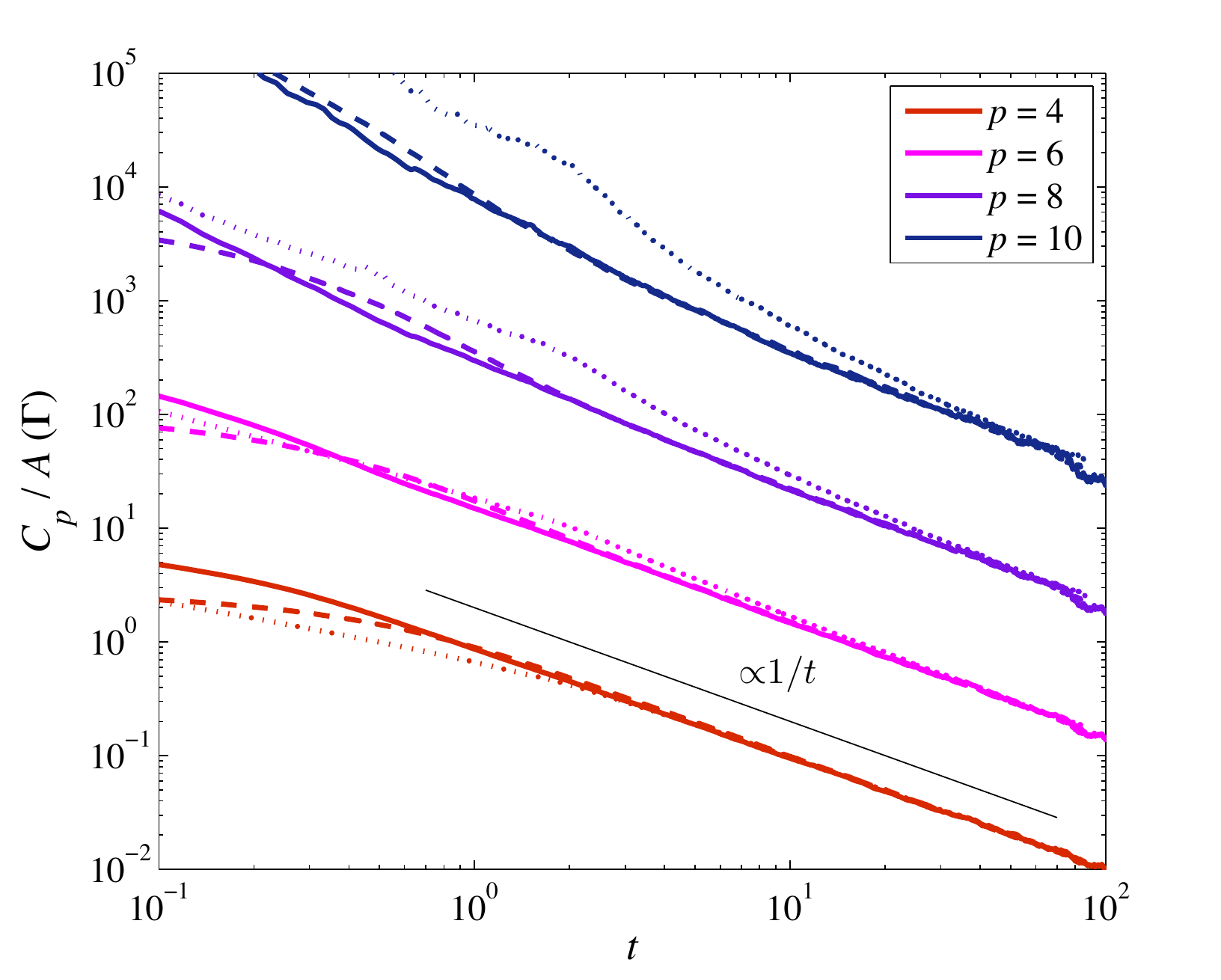}\label{fig:diffuse3_b}}
\hspace{-18pt}
\subfigure[]{\includegraphics[width=0.35\columnwidth]{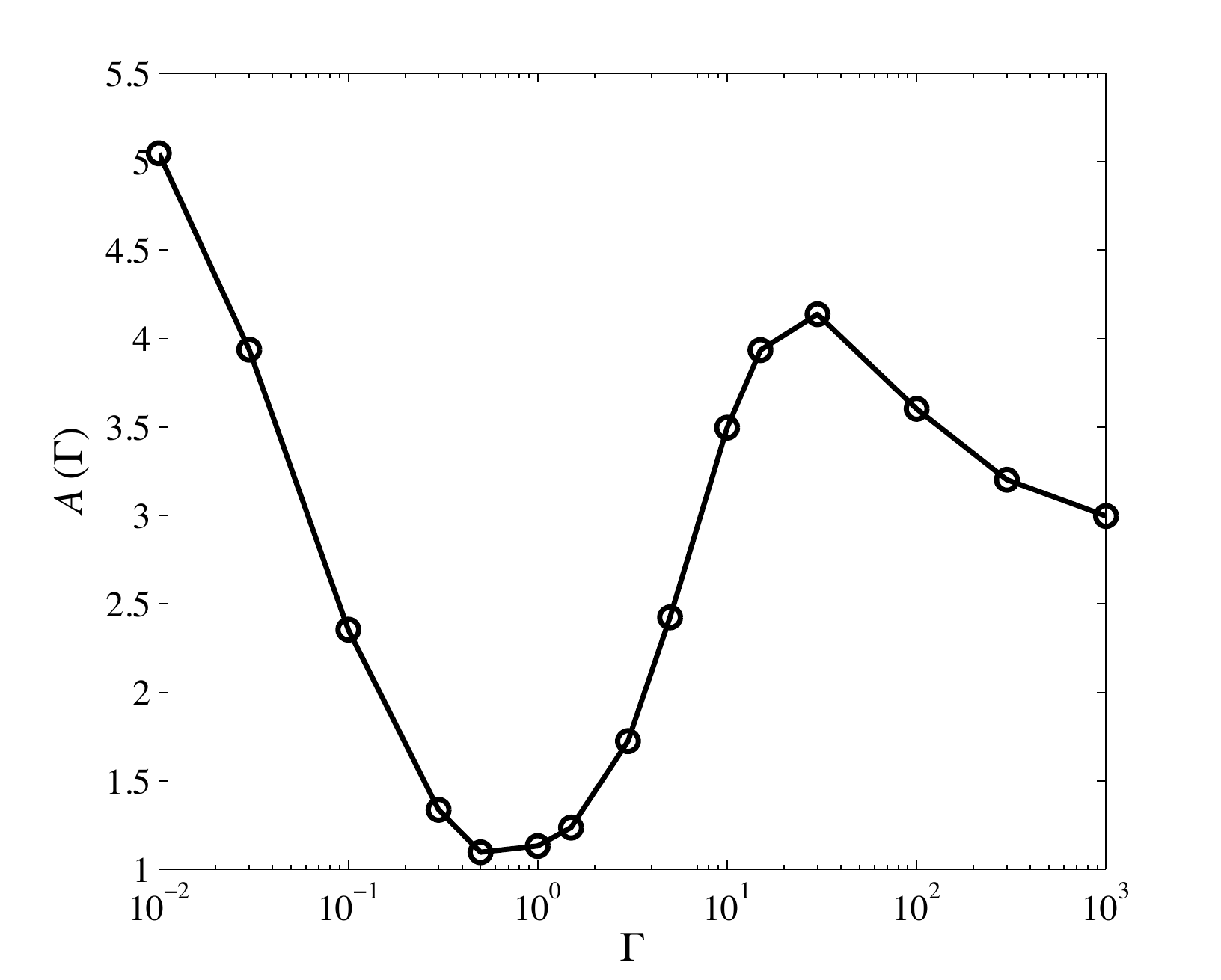}\label{fig:diffuse3_c}}
\caption{\label{fig:diffuse3} (a) PDF of $(X(t)-X(0))/(L/2)$ for different values of $\Gamma$ taken at $t^*$ such that $\langle\langle (X(t^*)-X(0))^2\rangle\rangle=(L/2)^2=\pi^2$. (b) Temporal evolution of normalized moments $C_p.$ defined by \eqref{Ec:Cp} for $\Gamma=0.1$ (solid lines), $\Gamma=1$ (dashed lines), and $\Gamma=10$ (dotted lines). (c) $A(\Gamma)$ showing the  dependence of $C_p$ on $\Gamma$, see equation \eqref{Eq:Cpnumresults}.}
\end{figure}

The normalized probability distribution function (PDF) of the displacements are shown on figure \ref{fig:diffuse2} for different times and different values of $\Gamma$. 
At early times and for low values of $\Gamma$, the PDF's present exponential tails, that are also observed at intermediate times for larger values of $\Gamma$. In all cases,  the PDF's approach a Gaussian distribution at very large times. Note that for sufficiently large values of $\Gamma$, some oscillations can be observed on the PDF tails at intermediate times. As we will now see, this is due to trapping by the zeros of $\sigma(x,t)$.
To emphasize this point, we compare the PDF of the displacement for different values of $\Gamma$ chosen at the time $t^*$ such that $\langle\langle (X(t^*)-X(0))^2\rangle\rangle=(L/2)^2=\pi^2$, i.e. the time when the typical distance traveled by the particles reaches a length of the order of the separation between two zeros (of the order of $L/2$ for smooth environments). The PDF of $(X(t^*)-X(0))/(L/2)$ is displayed on figure \ref{fig:diffuse3_a} for various values of $\Gamma$. It is apparent that the bumps appear at multiples of $L/2$ for large $\Gamma$'s.  With a high probability, a particle is initially located close to a zero of $\sigma$. When it travels away, there is a large chance that it stops in the neighboring zero, leading to a quasi multimodal distribution. 

To better quantify the large-time behavior of the displacement, we compute its higher-order moments and compare them to those corresponding to a Gaussian distribution. We thus define for even $p$
\begin{equation}
C_p=\frac{\langle (X(t)-X(0))^p\rangle}{\langle (X(t)-X(0))^2\rangle^{p/2}}-\frac{2^{p/2}}{\sqrt{\pi}}\Gamma\left[\frac{p+1}{2}\right],\label{Ec:Cp}
\end{equation}
where $\Gamma[z]$ denotes here the Gamma function. $C_p=0$ for all $p\geq2$ if $X(t)-X(0)$ is a Gaussian variable. $C_2=0$ by construction and $C_4$ is the Kurtosis. $C_p$ can be interpreted as a $p$-th order deviation from a Gaussian distribution. Data suggest that all the $C_p$'s display a $1/t$ behavior at large times. Its dependence on the amplitude $\Gamma$ and on the order $p$ is well represented by the functional form
\begin{equation}
C_p\simeq\frac{A(\Gamma)\lambda^{p/2}}{t}.\label{Eq:Cpnumresults}
\end{equation}
The temporal evolution of $C_p/A(\Gamma)$ for different values of $\Gamma$ and $p$, and the function $A(\Gamma)$ are displayed on figures \ref{fig:diffuse3_b} and \ref{fig:diffuse3_c}, respectively. As seen from \ref{fig:diffuse3_b}, the data for different $\Gamma$ and $p$ fixed collapse at large times. The limiting lines associated to different orders are equally space, confirming the dependence in $\lambda^{p/2}$ of the constant. The value of $\lambda$ was found to be $\lambda\approx15$.

\begin{figure}[h!]
\centering
\hspace*{-4pt}
\subfigure[]{\includegraphics[width=0.35\columnwidth]{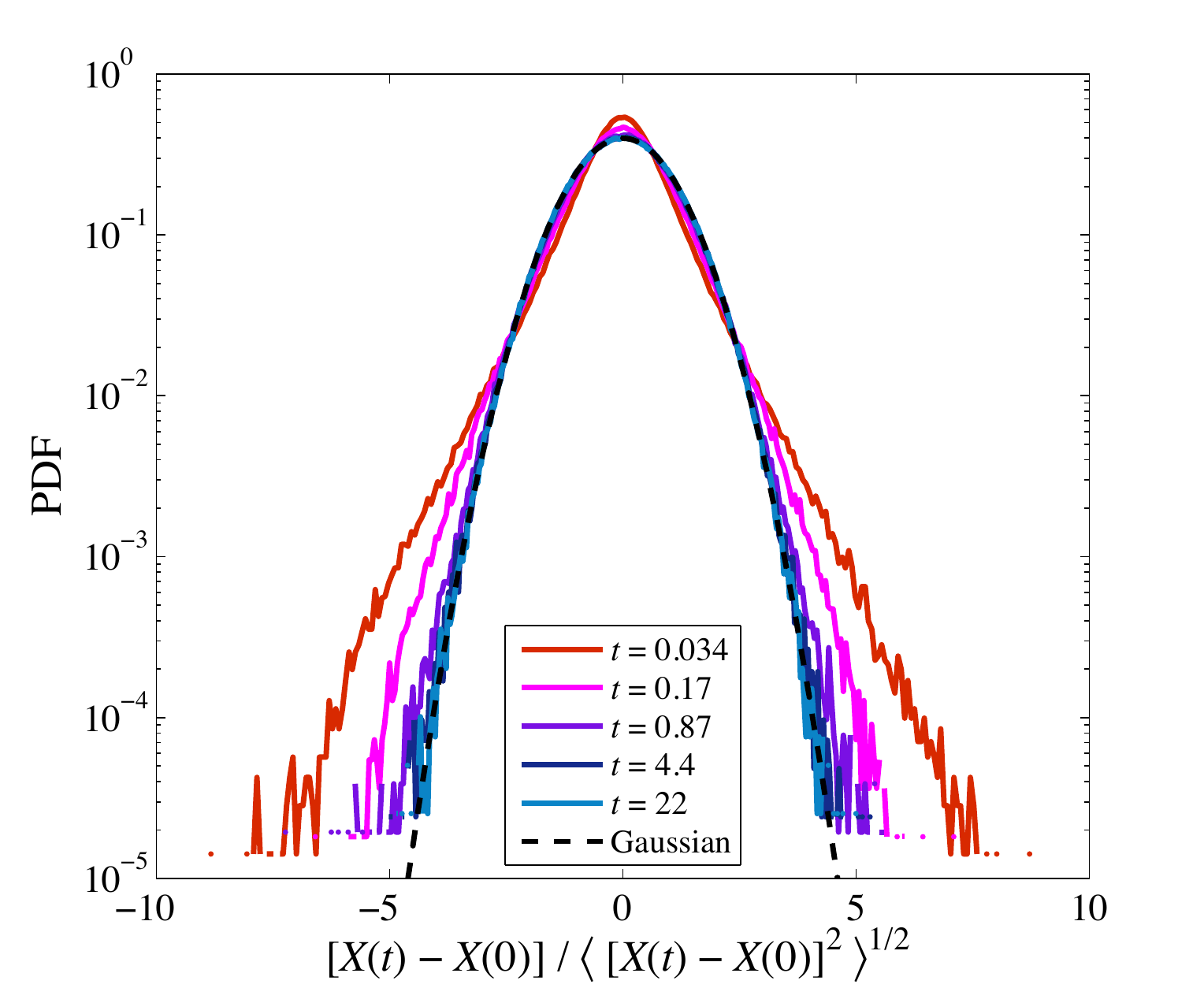}\label{fig:pdfDiffnonsmooth_a}}
\hspace{-18pt}
\subfigure[]{\includegraphics[width=0.35\columnwidth]{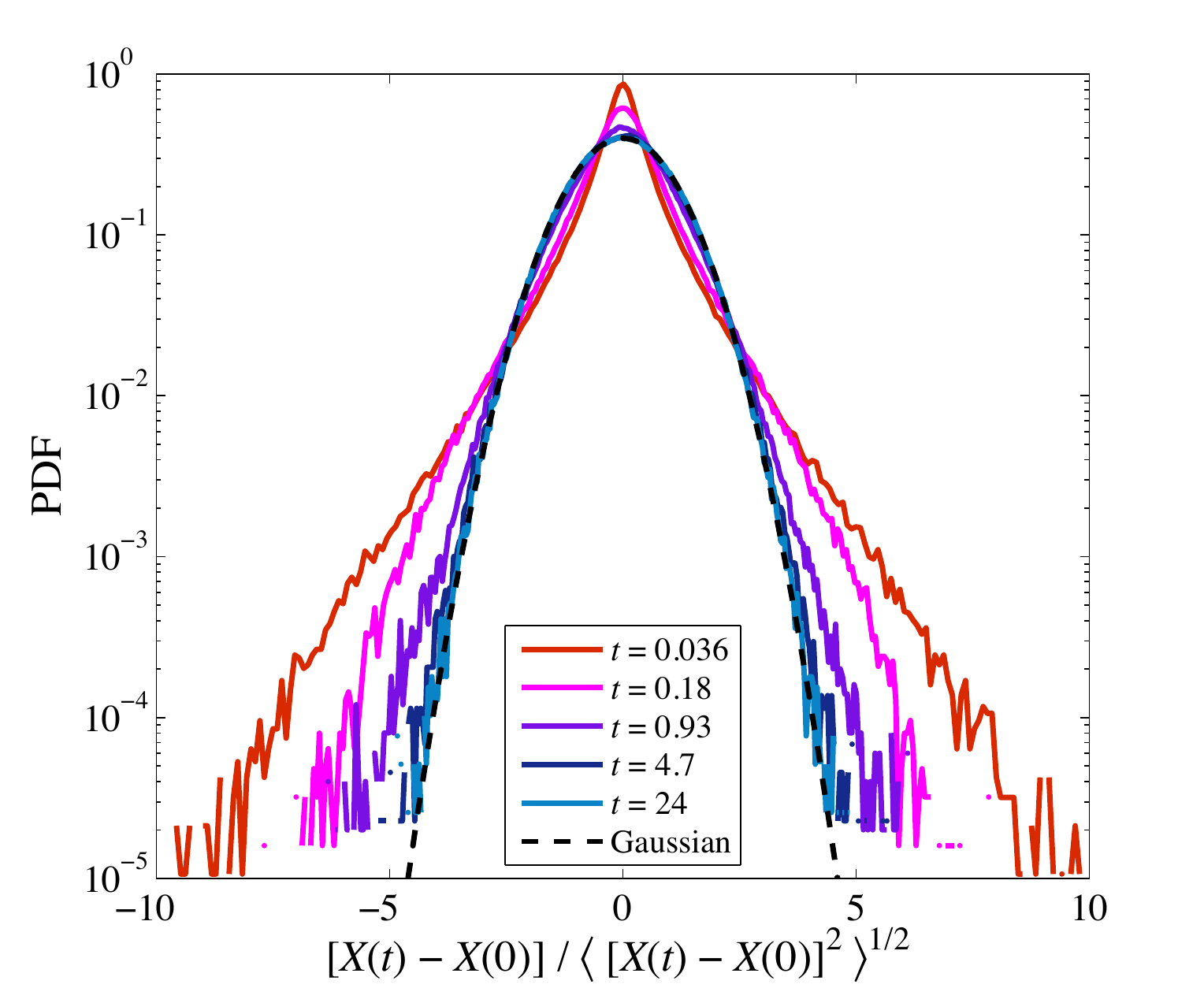}\label{fig:pdfDiffnonsmooth_b}}
\hspace{-18pt}
\subfigure[]{\includegraphics[width=0.33\columnwidth]{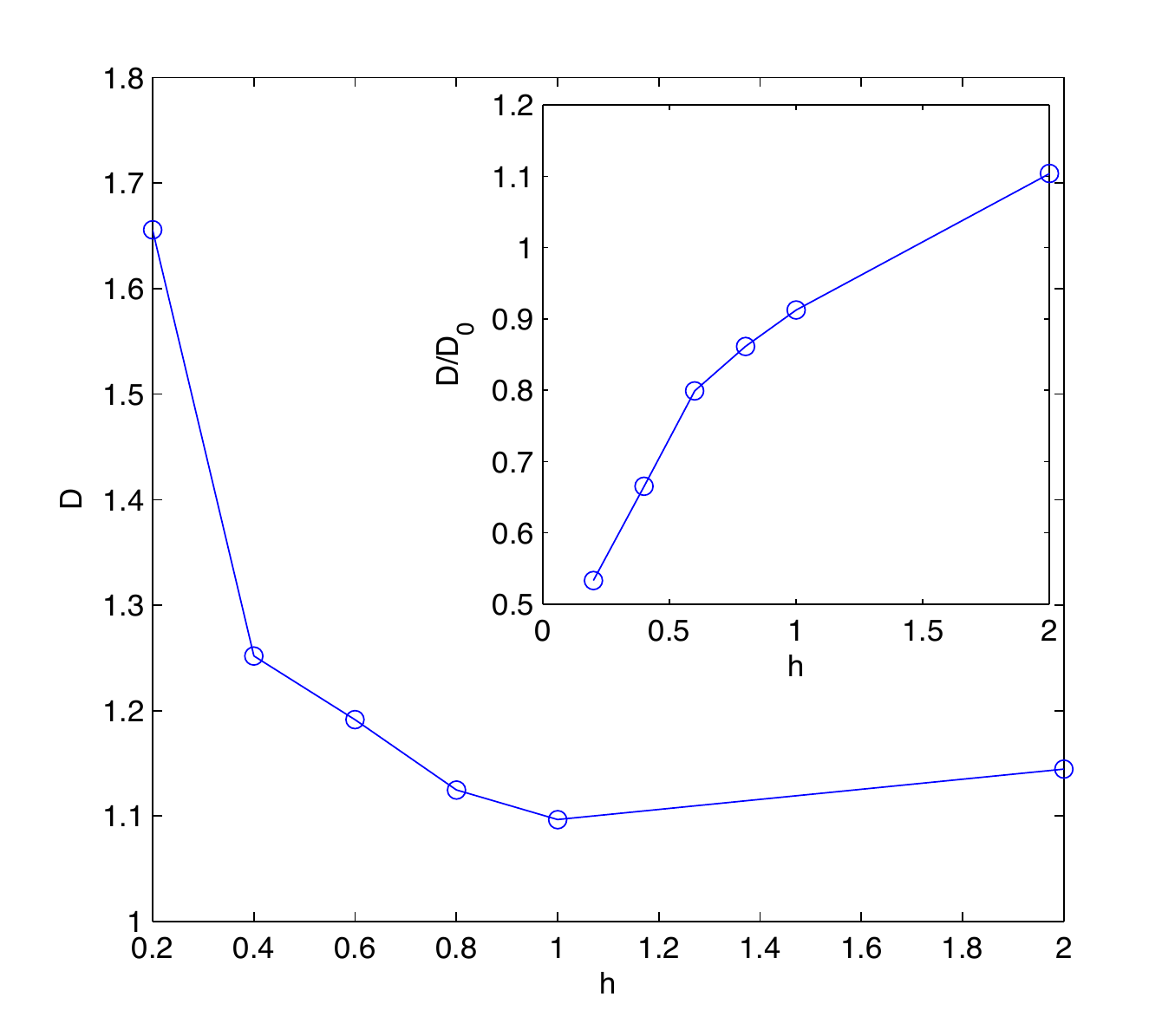}\label{fig:pdfDiffnonsmooth_c}}
  \caption{\label{fig:pdfDiffnonsmooth} Rescaled PDF of $X(t)-X(0)$ for
     $\Gamma=1$ and with (a) $h=0.2$ and (b) $h=0.6$ for different times. Dashed lines correspond to the Gaussian distribution. (c) Diffusion constant for $\Gamma=1$ as a function of $h$. The inset show the same data normalized by $D_0=\overline{\| \sigma\|_{L^2}^2}$. }
\end{figure}
Different simulations have been performed for various values of the exponent $h$. A behavior similar to the smooth case was found: at large times particles diffuse and their statistics become Gaussian. The rescaled PDF's of the displacement are shown on figure \ref{fig:pdfDiffnonsmooth_a} and \ref{fig:pdfDiffnonsmooth_b} for $\Gamma=1$ and for $h=0.2$ and $h=0.6$, respectively.
At large times $\overline{\langle (X(t)-X(0))^2\rangle}$ also present a linear grow consistent with equation \eqref{Eq:diffLin} (data not shown). The measured diffusion constant $D(\Gamma,h)$ is displayed on figure \ref{fig:pdfDiffnonsmooth_c} for $\Gamma=1$ and different values of $h$. The apparent divergence at small values of $h$ is due to the divergence of the $L^2$ norm of $\sigma(x,t)$ for $h\to0$. The inset of figure \ref{fig:diffuse3_c} shows the dependence of the diffusion constant normalized by $D_0=\overline{\| \sigma\|_{L^2}^2} = \zeta(1+2h)$. It is seen that the importance of zeros on the diffusion constant increases when $h$ decreases.

\section{Density fluctuations\label{Sec:DensityFluct}}

So far, we have studied the diffusion properties of particles described by the stochastic equation \eqref{eq:sde}. We now turn to studying the statistical properties of the mass density $\rho$ that evolves according to the diffusion-like equation \eqref{Eq:ModelD}. As we have previously observed, the spatial fluctuations of $\rho$ are correlated with the distribution of zeros of the random environment $\sigma({\bf x},t)$. More precisely, we observe from equation \eqref{Eq:ModelD} that the flux of mass at ${\bf x}$ is directly given by
\begin{equation}\label{Eq:fluxmass}
{\bf J}({\bf x},t)=\Gamma\,\sigma({\bf x},t)\left[2{\bf \nabla}\sigma({\bf x},t)\rho({\bf x},t)+\sigma({\bf x},t){\bf \nabla}\rho({\bf x},t)\right].
\end{equation}
The flux hence vanishes at the zeros of ${\sigma}({\bf x},t)$, so that they act like sinks and concentrate all of mass if they were not moving. However, due to their diffusion, the zeros are not able to indefinitely concentrate mass and the density saturates in their neighborhood. In parallel, as the total mass is conserved, this concentration process leads to the creation of voids in the regions where $\sigma$ is order one.

In the following we consider the case $D=1$. Under some assumptions of homogeneity and ergodicity of the dynamics, the PDF of the mass density can be written as
\begin{eqnarray}
P(\bar{\rho})&=&\frac{\mathrm{d}}{\mathrm{d}\bar{\rho}}\left({\rm measure}(\{(x,t)\,|\, \rho(x,t)\le\bar{\rho}\})\right)\\
  &=&\lim_{T\to\infty}\frac{1}{2\pi T}\int\limits_0^T \int\limits_0^{2\pi} \delta\left({\rho(x,t)-\bar{\rho}}\right)\mathrm{d}x\,\mathrm{d}t\equiv\lim_{T\to\infty}P_T(\bar{\rho})\label{Eq:defP}
\end{eqnarray}
This will be the main quantity studied in the next sections. As we will see it strongly depends on the parameters $\Gamma$ and $h$. The definition (\ref{Eq:defP}) in terms of a space-time average will be particularly useful to attack first the stationary case and then to develop phenomenological arguments on its behavior in the non-stationary case.

\subsection{Smooth random environments}

\begin{figure}[h!]
\centering
\subfigure[]{\includegraphics[width=.45\columnwidth]{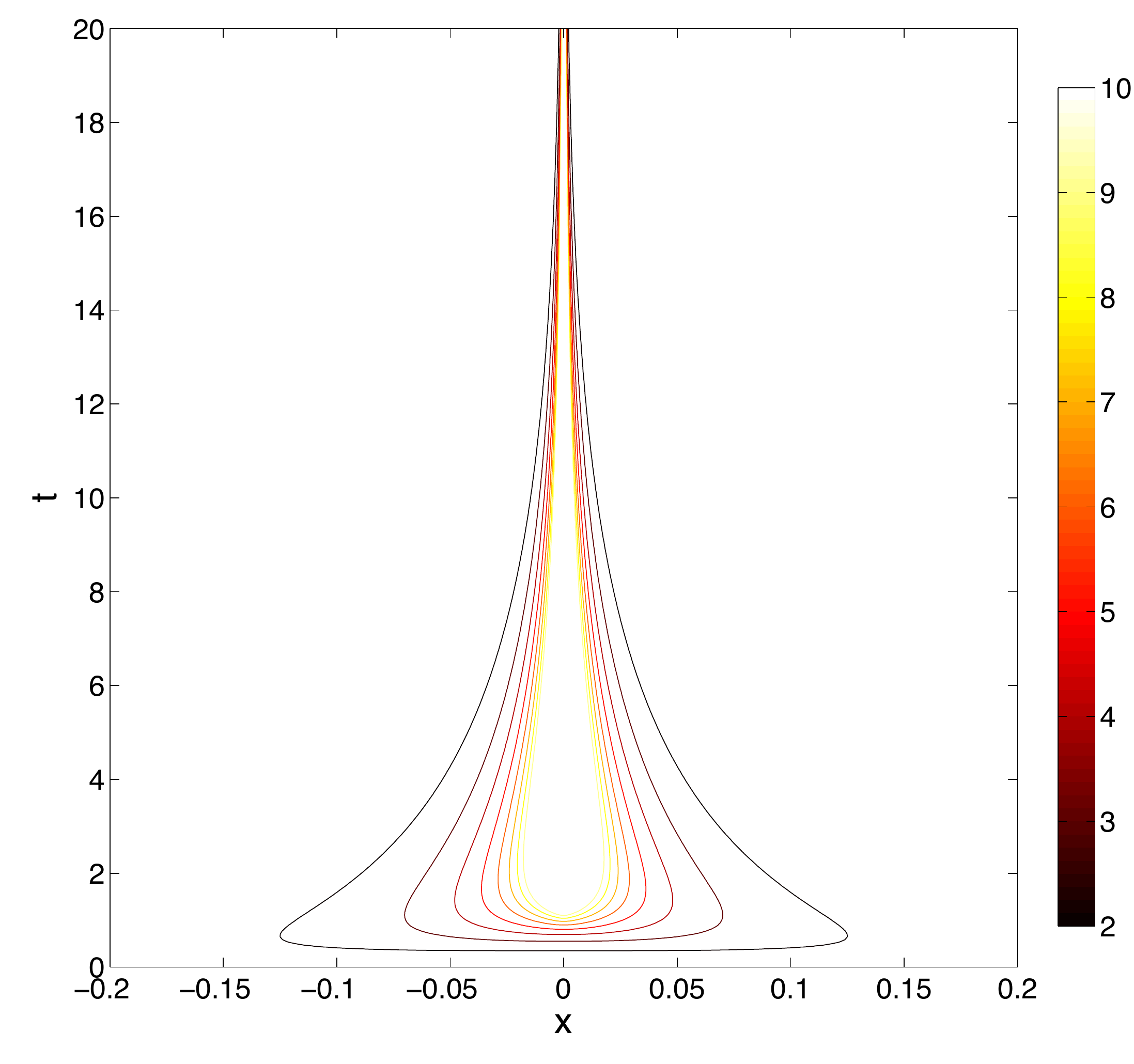}\label{Fig:ContourRho_a}}
\quad
\subfigure[]{\includegraphics[width=0.45\textwidth]{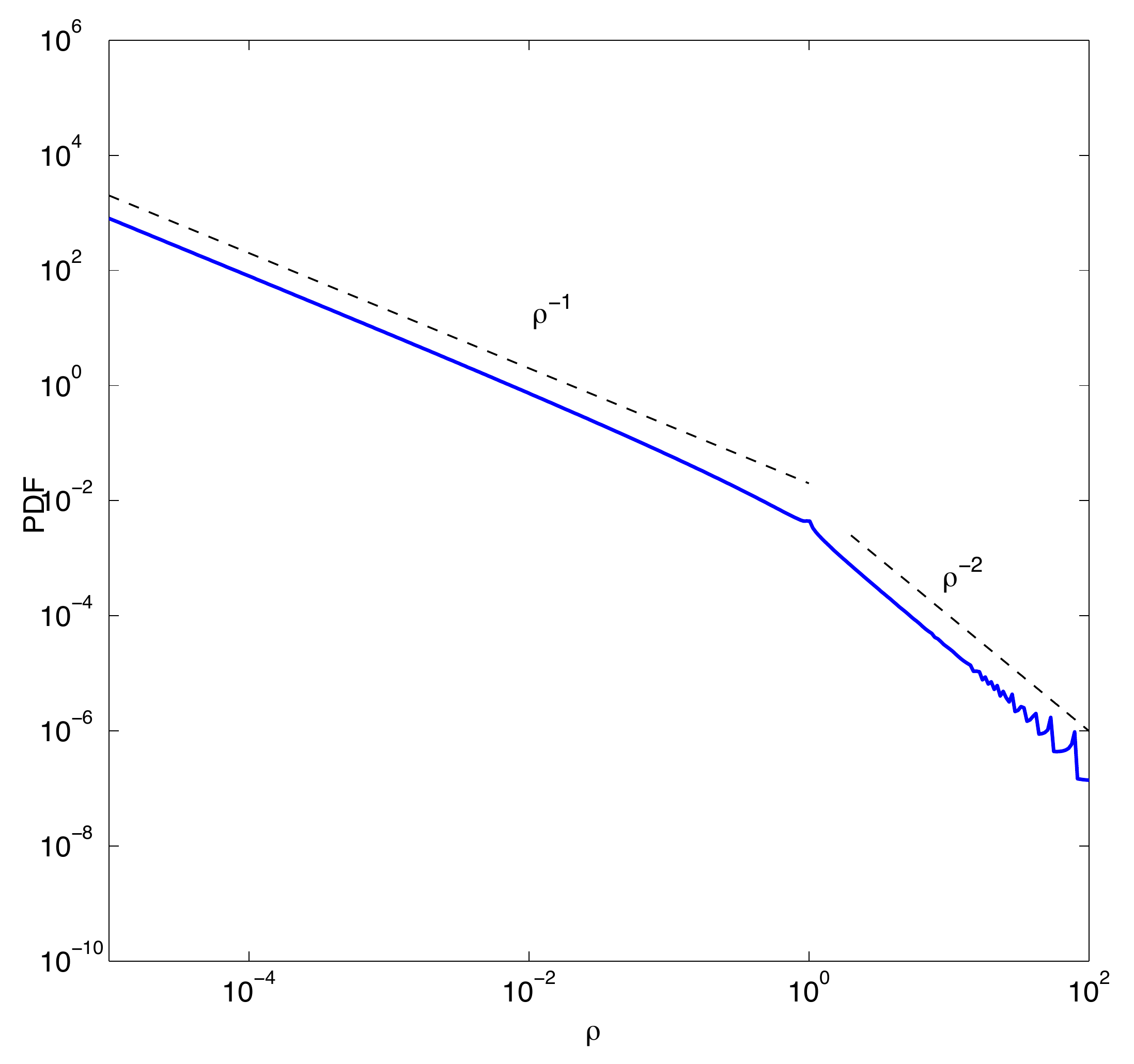}\label{Fig:ContourRho_b}}
  \caption{\label{Fig:ContourRho} (a) Contour lines of the analytic solution $\rho(x,t)$ explicitly given in equation \eqref{Eq:RhoStatSol}. (b) Probability distribution $P(\rho)$ \eqref{Eq:defP}. Dashed lines display the scalings of equations \eqref{Eq:predSN} and \eqref{Eq:predlowmass}.}
\end{figure}

Let us first consider the limit $\mathrm{Ku}\to\infty$, i.e. when  the ejection $\Gamma$ is infinitely larger than the inverse of the diffusion time of the zeros of $\sigma$. Note that, in the limit of a time-independent environment, no stationary state can be achieved. Mass will then concentrate in the zeros and at $t=\infty$ the density will become atomic, supported on these points. For sufficiently large but finite times, as almost all the mass is concentrated around the zeros of $\sigma$, the ejection rate can be Taylor-expanded in the vicinity of these points. Without loss of generality, let us assume that a zero appears at the origin $x=0$ at $t=0$ and does not move. In this limit, equation \eqref{Eq:ModelD} reduces to
\begin{equation}
  \partial_t \rho =\Gamma C^2 \frac{\partial^2}{\partial{x}^2} (x^{2} \rho),\label{Eq:modelstat}
\end{equation}
where $C = \mathrm{d}\sigma/\mathrm{d}x|_{x=0}$. Rescaling time allows us to set $\Gamma\,C^2=1$. Without loss of generality we moreover consider an infinite domain and an initial condition with compact support: $\rho(x,0) = 1$ for $|x| \le 1$ and $0$ elsewhere. Equation \eqref{Eq:modelstat} can be analytically integrated by using the method of characteristics. For $x>0$ the change of variable $u(y,t)=e^{2 t}\rho(x=e^{y-3 t},t)$ leads to the homogeneous heat equation that can be easily solved by using the corresponding Green function. The solution for $x<0$ is obtained by symmetry. One then obtains
\begin{equation}
  \rho(x,t) =\frac{\mathrm{e}^{2t}}{\sqrt{4\pi t}}\int_0^\infty \exp\left[{-\frac{(z -\log{|x|}-3 t)^2}{4 t}}\right] \mathrm{d}z= \frac{1}{2}\mathrm{e}^{2t}\mathrm{erfc}\left[\frac{\ln
    |x|+3t}{2\sqrt{t}}\right],\label{Eq:RhoStatSol}
\end{equation}
where $\mathrm{erfc}(z)=\frac{2}{\sqrt{\pi}}\int_{z}^\infty \mathrm{e}^{-s^2}\mathrm{d}s$ is the complementary error function. The time-averaged distribution $P_T(\bar{\rho})$ defined in equation \eqref{Eq:defP} can be rewritten as
\begin{equation}
P_T(\bar{\rho})= \frac{1}{2\pi T}\int\limits_0^T \int\limits_0^{2\pi}\frac{\delta(x-x_{\bar{\rho}}(t))}{|\pder{\rho(x(t),t)}{x}|}\mathrm{d}x\,\mathrm{d}t,\label{Eq:Pdfbis}
\end{equation}
where $x_{\bar{\rho}}(t)$ is such that $\rho(x_{\bar{\rho}}(t),t)=\bar{\rho}$. It exists only for $t>\ln{\bar{\rho}}/2$ . The contour lines of $\rho(x,t)$ are displayed on figure \ref{Fig:ContourRho_a}.
The high-density zones are concentrated in a narrow region of space. This will justify the saddle-point approximation used below. With the change of variables $t \mapsto \lambda=2t/\ln{\bar{\rho}}$, introducing $\mu(\lambda;\bar{\rho}) = -\ln x_{\bar{\rho}}(t) / \ln \bar{\rho}$, and after some algebra, equation \eqref{Eq:Pdfbis} becomes
\begin{equation}
P_T(\bar{\rho})=\frac{1}{T}\sqrt{\frac{\ln^3{\bar{\rho}}}{4\pi}}\int\limits_1^\frac{2T}{\ln{\bar{\rho}}}\sqrt{\lambda}e^{-\ln{\bar{\rho}}F(\lambda;\bar{\rho})}\mathrm{d}\lambda,\label{Eq:PDFrhoAsym}
\end{equation}
with
\begin{eqnarray}
F(\lambda;\bar{\rho})&=& \lambda +\mu(\lambda;\bar{\rho})-\frac{1}{2\lambda}\left(\frac{3}{2}\lambda-\mu(\lambda;\bar{\rho})\right)^2\\
\nonumber\mu(\lambda;\bar{\rho})&=&\frac{3}{2}\lambda -\sqrt{\frac{2\lambda}{\ln{\bar{\rho}}}}\mathrm{erfc}^{-1}\left(\frac{2}{\bar{\rho}^{\lambda-1}}\right)\approx\frac{3}{2}\lambda -\sqrt{2\lambda(\lambda-1)},
\end{eqnarray}
where in the last equation the asymptotic of $\mathrm{erfc}^{-1}(z)\approx\sqrt{\log{(1/z)}}$ for $z\approx 0$ has been used (recall that $\lambda>1$). Using a saddle-point approximation
for $\ln{\bar{\rho}}\gg1$ of the integral in equation \eqref{Eq:PDFrhoAsym}, we obtain at leading order (dropping the bar over $\rho$)
\begin{equation}
P_T(\rho)\sim \frac{1}{\rho^{\alpha}}, \mbox{ with } \alpha = \inf_\lambda F(\lambda) = 2.\label{Eq:predSN}
\end{equation}
This scaling is clearly observed on figure \ref{Fig:ContourRho_b}, where equation \eqref{Eq:ModelD} is integrated using the stationary environment $\sigma(x,t)=\sin{x}$.

Note that a clear power-law scaling is also seen on figure \ref{Fig:ContourRho}.b at small values of the mass density. To explain this scaling, let us to consider a cell of length $\Delta x$ located at $x_0$ far away from a zero of $\sigma(x)$ and where the density reaches its minimum. At leading order the mass ejected from this cell between $t$ and $t+dt$ is proportional to $J\simeq\rho(x,t)\Gamma\partial_x\sigma^2(x)|_{x=x_0}$ as $\partial_x\rho\approx0$ at this point. Considering that $\partial_x\sigma^2(x)$ is constant near $x_0$ yields an exponential decay of the mass in the cell. Introducing in equation \eqref{Eq:defP} a density of mass with an exponential decay $\rho(x,t)\sim\rho_0e^{-\sigma_0^2 t}$ (with $\sigma_0^2$ an effective rate) directly leads to
\begin{equation}
P_T(\rho)\sim \frac{1}{\rho},\label{Eq:predlowmass}
\end{equation}
which is in agreement with figure \ref{Fig:ContourRho}.b.

\begin{figure}[h!]
\centering\includegraphics[width=.8\columnwidth]{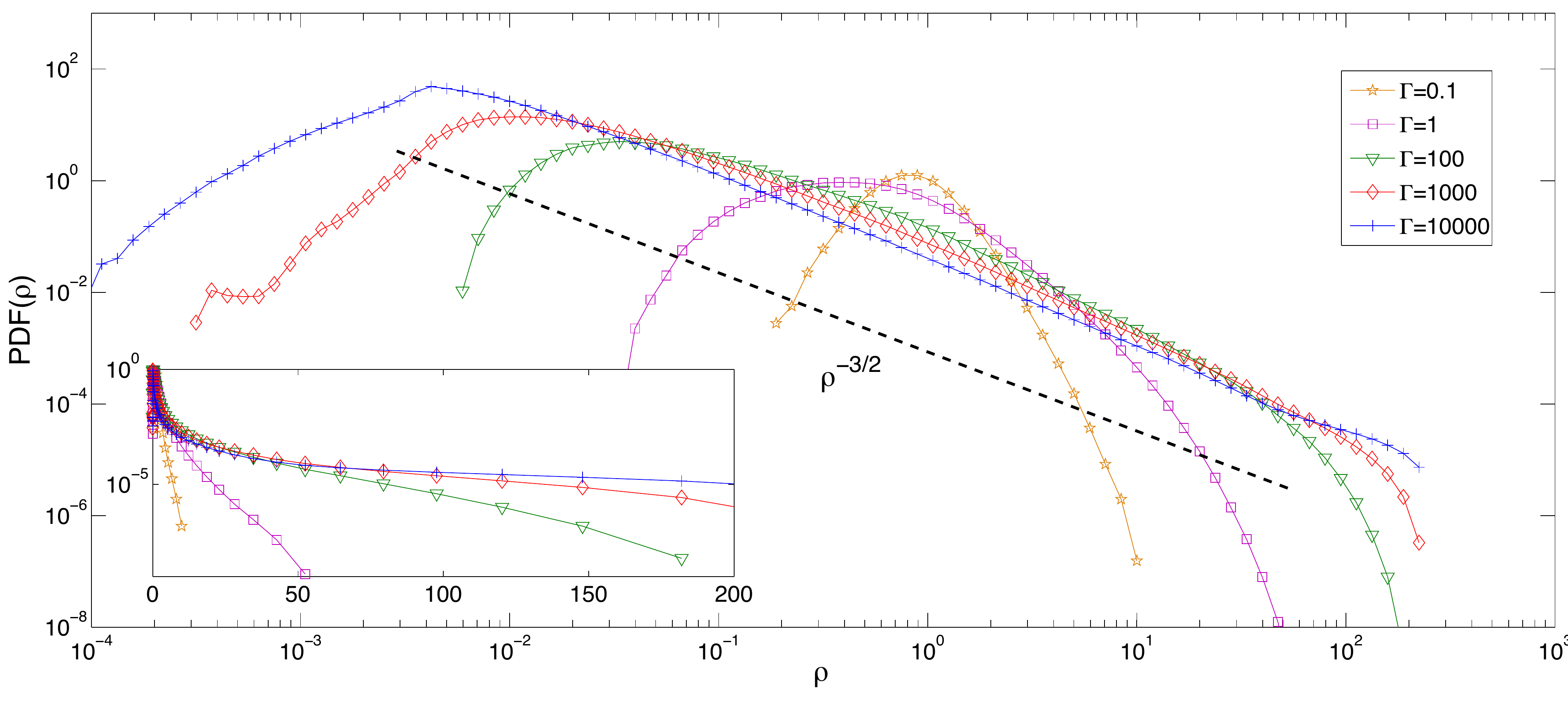} 
  \caption{\label{fig:PDFTwomodes} Log-log plot of the PDF of the density for $K=1$ and different values of $\Gamma$. Inset: Same figure in log-lin.
}
\end{figure}

\begin{figure}[h!]
\centering\includegraphics[width=.7\columnwidth]{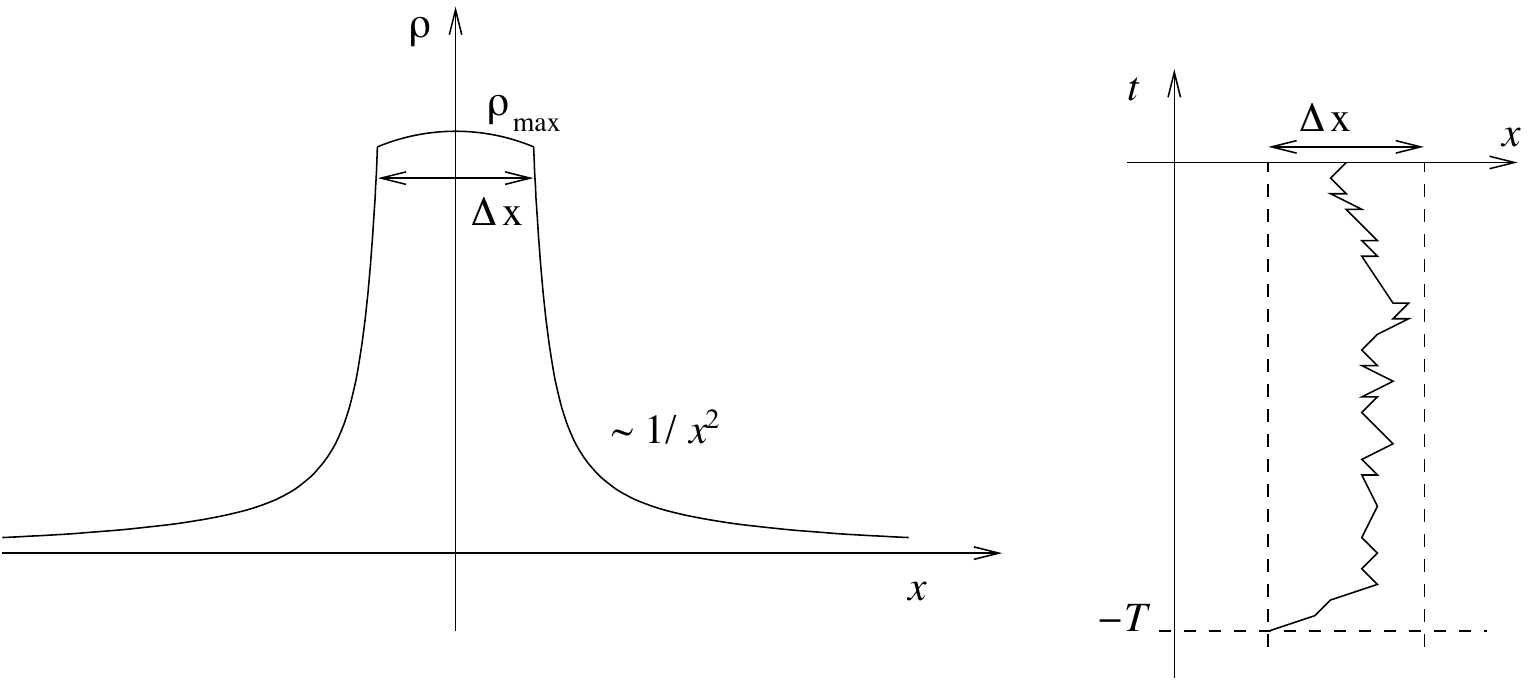} 
  \caption{\label{fig:sketch_density} Left: sketch of the density field in the vicinity of a zero of the ejection rate $\sigma^2$ located at $x=0$; the density grows as $x^{-2}$ and saturates to $\rho_\mathrm{max}$ at a distance $\Delta x\sim \rho_\mathrm{max}^{-1/2}$ of the zero. Right: sketch of the diffusive time evolution of a zero of $\sigma^2$, which has remained at a distance less than $\Delta x$ for a time $T$ before the reference time $t=0$.
}
\end{figure}

As we will now see the situation is rather different in the non-stationary case where the motion of the zeros of the ejection rate limits the process of mass concentration. Now, the coefficients $\sigma_r(t)$ and $\sigma_i(t)$ of equation \eqref{Eq:Sigma1} are the Ornstein--Uhlenbeck processes defined in equations \eqref{Eq:defSigma} and \eqref{Eq:AmpSigma}. The PDF of $\rho$ for different values of $\Gamma$ is presented on figure \ref{fig:PDFTwomodes}.
The time needed for the density to accumulate near the zero of the ejection rate $\sigma^2$ is of the order of $\Gamma^{-1}$. Hence, for small values of $\Gamma$ (i.e.\ $\mathrm{Ku}\ll1$), the fast temporal variations of $\sigma(x,t)$ do not leave enough time for the system to cumulate mass. It is apparent in this case that most of mass is distributed around the mean mass $\langle \rho\rangle=1$ (see the $\Gamma=0.1$ curve represented by orange pentagrams). However, for large but finite values of $\mathrm{Ku}$, accumulation is fast enough and a resulting $\rho^{-3/2}$ scaling is observed. This scaling can be derived by considering that for large $\mathrm{Ku}$ the system rapidly relaxes to a quasi-stationary solution in a time of order $1/\Gamma$ and stays there for a time of the order of the correlation time of $\sigma$. Far enough from a zero, the environment behaves as $\sigma(x)\simeq C\,x$ and its temporal evolution can be neglected. The density converges there to a quasi-stationary solution, which is such that $\partial_x^2(\sigma^2\rho)\approx0$, so that $\rho\approx\sigma^{-2}\propto 1/x^{2}$.  Closer to the zero, the time variations of its location become faster than the concentration of mass and the density saturates to a finite value $\rho_\mathrm{max}$ (see the left-hand side of figure \ref{fig:sketch_density}). 
The transition between these two regimes occurs at a distance $\Delta x$ from the zero, which by continuity satisfies $\Delta x\sim \rho_\mathrm{max}^{-1/2}$. The length $\Delta x$ is of the order of the distance travelled by the zero during a timescale equal to that of mass concentration. Hence, if we assume that the zero diffuses, one has  $\Delta x\sim \Gamma^{-1/2}$, so that the typical value of $\rho_\mathrm{max}$ is of the order of $\Gamma$. The distribution of density has thus a crossover at $\rho\sim\Gamma$. When $\rho\ll\Gamma$, that is for values much below the plateau at $\rho_\mathrm{max}$, the behavior is dominated by the divergence of the quasi-stationary density profile. Introducing $\rho(x)\sim 1/x^2$ in equation \eqref{Eq:defP} straightforwardly leads to the algebraic behavior
\begin{equation}
P(\rho)\sim\rho^{-3/2} \mbox{ for } 1\ll\rho\ll\Gamma.
\end{equation}
Note that this value of the exponent in the large-density intermediate tail is very robust. It has also been observed for larger numbers of modes and with a environment of the form $\sigma(x,t)=\sqrt{\Gamma}\cos{(x-ct)}$ (data not shown).

The behavior at $\rho\gg\Gamma$ of the PDF of density is related to the large fluctuations of $\rho_\mathrm{max}$ and thus to the events when the zero does not move much during a time larger than $\Gamma^{-1}$. More precisely, if at a fixed time we observe a large value of $\rho_\mathrm{max}$, this means that the zero has not moved by a distance larger than $\rho_\mathrm{max}^{-1/2}$ during a time interval of length $T$ larger than $\Gamma^{-1}$ (see the right-hand side of figure \ref{fig:sketch_density}). When the zeros diffuse, this probability relates to the first exit time distribution of a diffusive process. The probability density of the first time $T$ at which the Wiener process exit the interval $|x|<\Delta x$ is $\sim \exp(-C T/\Delta x^2)$ (see, e.g., \cite{feller_1971}). We thus obtain 
\begin{equation}
\mathrm{Prob}\,(\rho_\mathrm{max}>\rho) \sim \int_{\Gamma^{-1}}^\infty \mathrm{e}^{-C\,T\,\rho} \,\mathrm{d}T.
\end{equation}
For $\rho\gg\Gamma$ the leading-order behavior is given by $T=\Gamma^{-1}$. This leads to the following exponential cutoff for the PDF of mass density
\begin{equation}
P(\rho) \sim \mathrm{e}^{-C\,\Gamma^{-1}\,\rho} \mbox{ for } \rho\gg\Gamma.
\end{equation}
This exponential behavior is confirmed numerically as can been seen on the lin-log plot showed in the inset of figure \ref{fig:PDFTwomodes}.

\subsection{Non-smooth random environment\label{Sec:NonSmooth}}

\begin{figure}[h!]
\centering
\subfigure[]{\includegraphics[width=0.45\columnwidth]{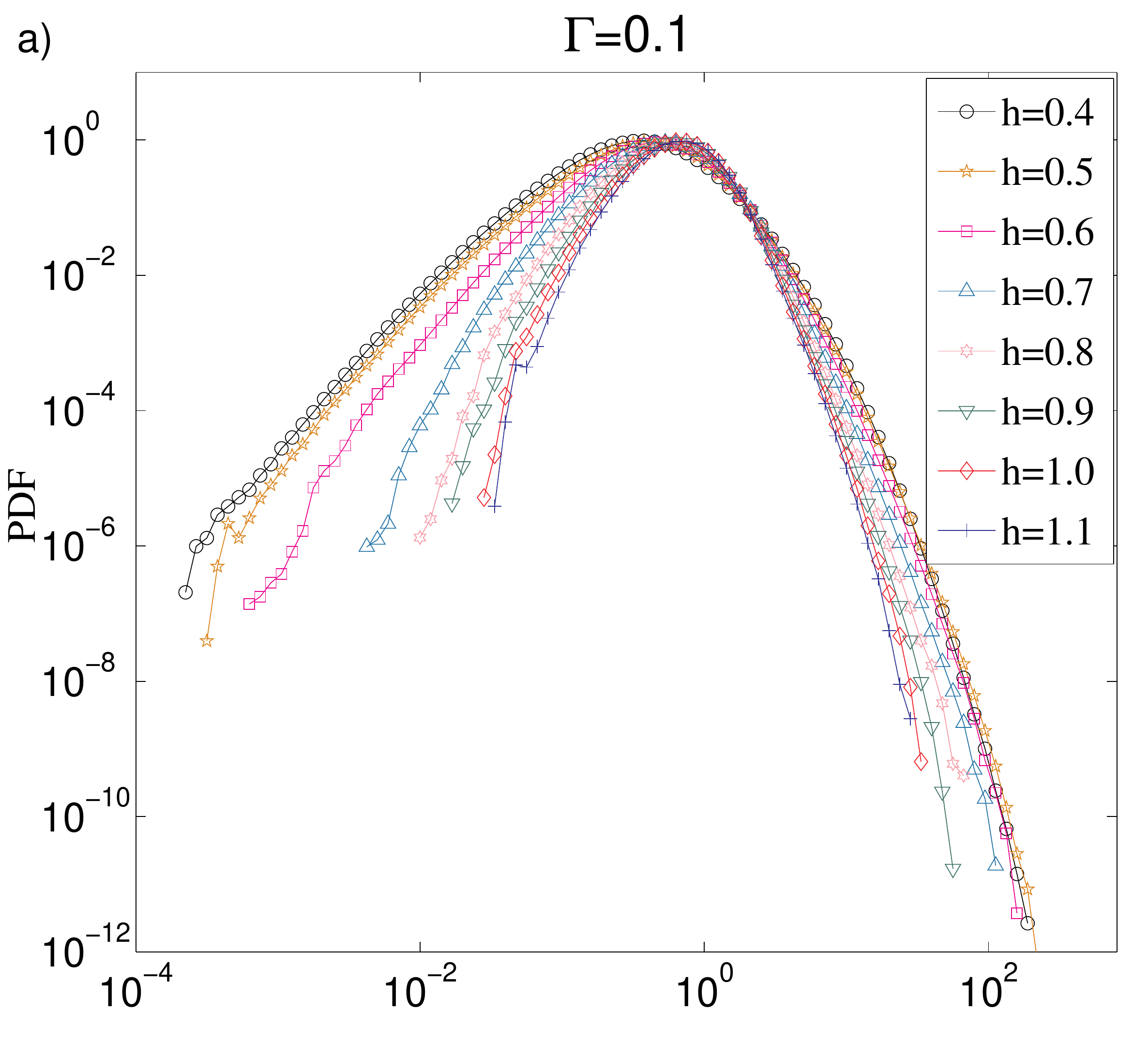}\label{Fig:PDFs_vs_h_a}}
\hfill
\subfigure[]{\includegraphics[width=0.45\columnwidth]{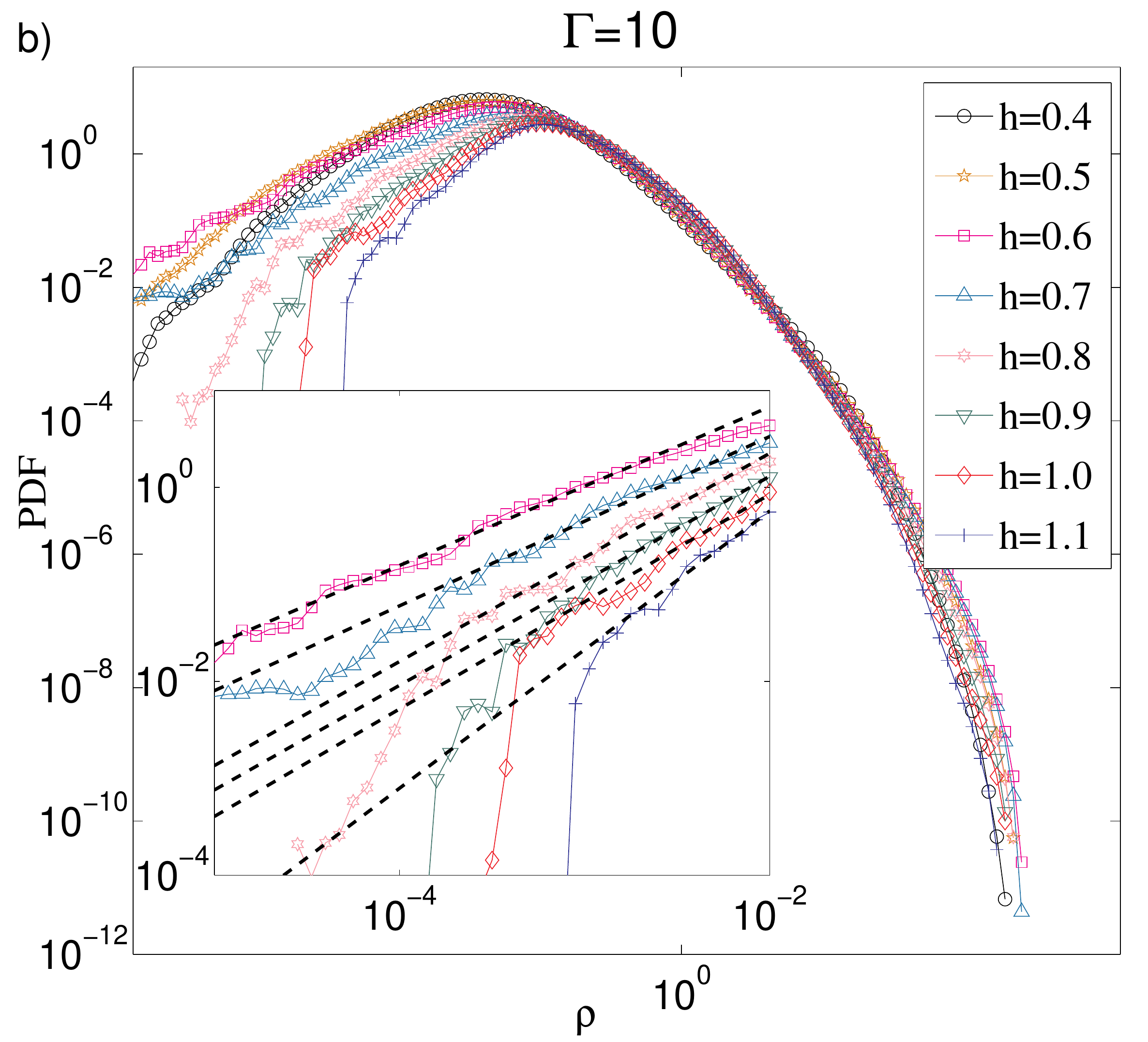}\label{Fig:PDFs_vs_h_b}}
  \caption{\label{Fig:PDFs_vs_h} Log-log plot of the PDF of the  density $\rho$ for different values of $h$ and (a) $\Gamma=0.1$ and (b) $\Gamma=10$. }
\end{figure}

\begin{figure}[h!]
\centering
\subfigure[]{\includegraphics[width=0.4\columnwidth]{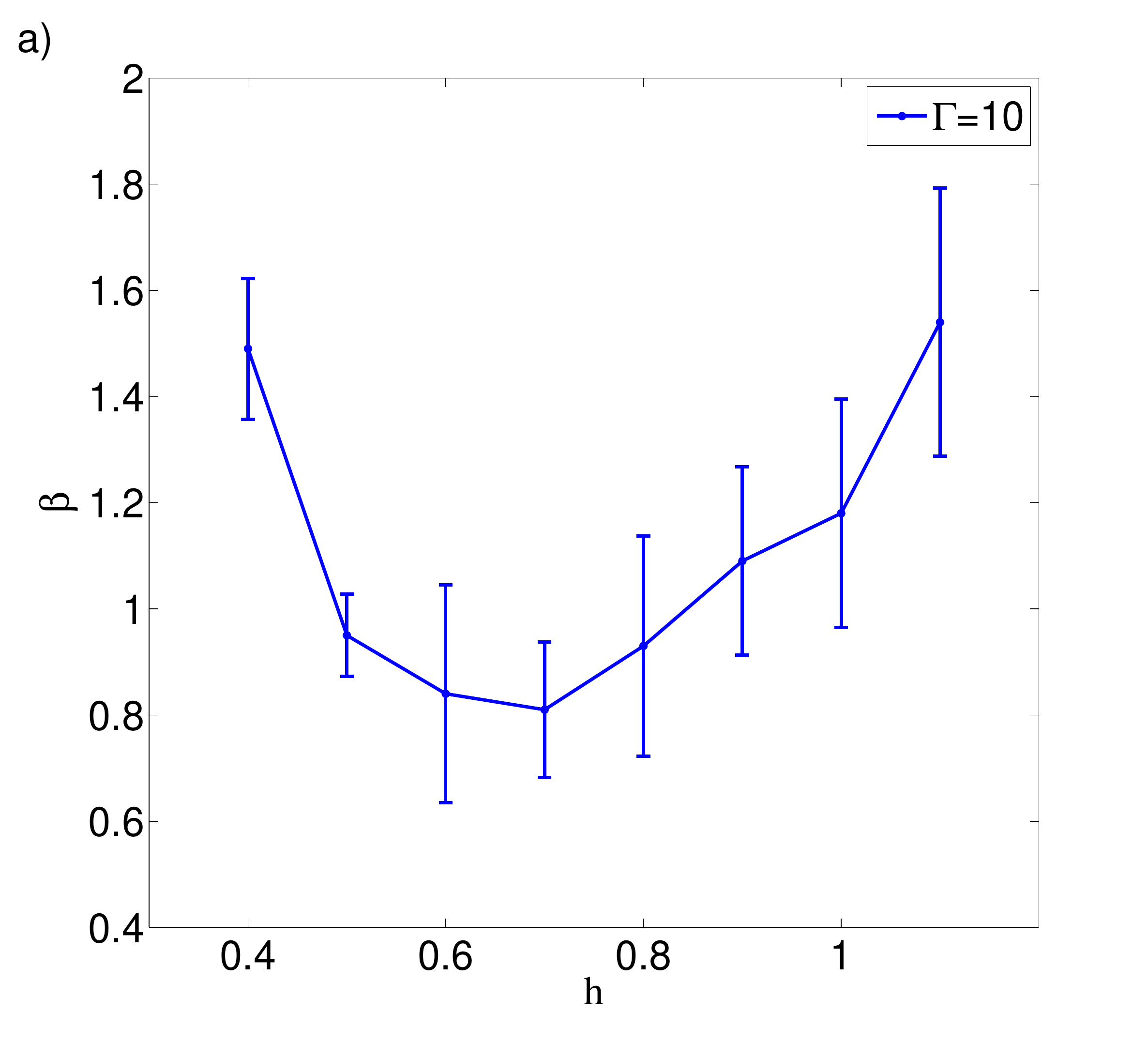}\label{FIG:exp&var_a}}
\qquad
\subfigure[]{\includegraphics[width=0.4\columnwidth]{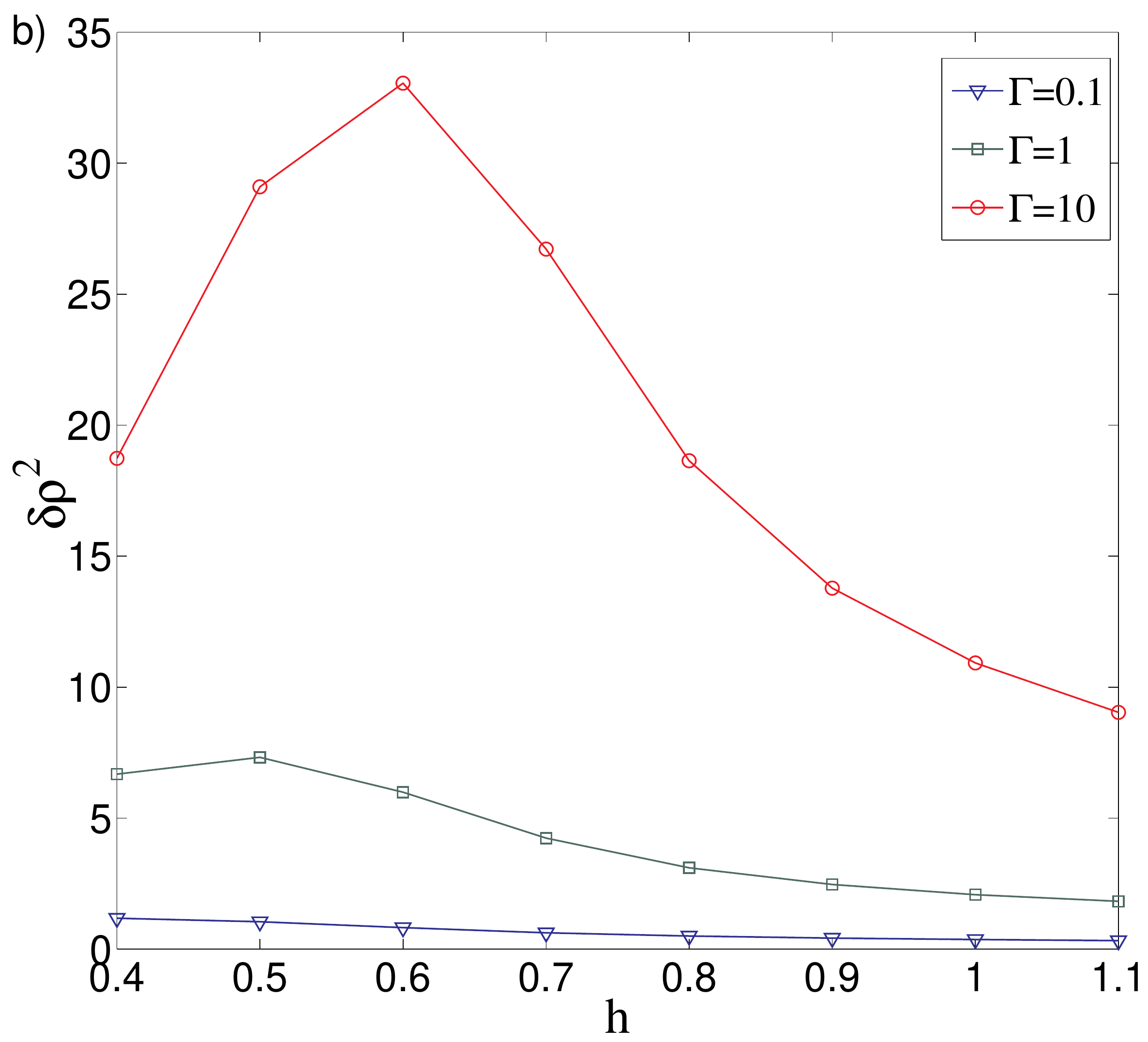}\label{FIG:exp&var_b}}
  \caption{\label{FIG:exp&var} (a) exponent $\beta$ of the power-law fit at the small-mass tail of the PDF's on \ref{Fig:PDFs_vs_h_b}, i.e. for $\Gamma=10$ and various $h$ as labeled. (b) Variance $\langle \delta\rho^2\rangle=\langle (\rho -\langle\rho\rangle)^2\rangle$ as function of $h$ for different values of $\Gamma$.
  }
\end{figure}

We now turn to the case of a time-dependent non-smooth ejection rate $\sigma$. Figure \ref{Fig:PDFs_vs_h} presents the PDF of $\rho$ for a number of runs with different values of $\Gamma$ and $h$.
Remark that for large $\Gamma$ a power-law behavior is observed at small masses ---\,see inset of figure \ref{Fig:PDFs_vs_h_b}. These tails can be understood using a phenomenological argument similar to the one used in \cite{Bec:NJP2007}. Consider an extreme event leading to a very low density in a cell where only ejection of mass has taken place for some time. For such a configuration we expect an exponential decay of the density and thus a time of order $T\propto-\Gamma^{-1} \ln{\rho}$ to reach a small density $\rho$ (where $\sigma_0$ is an effective rate).  The probability of having this configuration depends only on the properties of the environment. Let $p<1$ be the probability of having such a configuration. As the environment de-correlates in a time order one, the number of independent realizations of the environment that is needed for the full process to take place during a time $T$, is $\propto T \propto -\Gamma^{-1} \ln{\rho}$. Therefore, the probability of this complete event is $p^{-C\ln\rho/\Gamma}$, where $C$ is a proportionality constant that depends on $h$. We thus obtain that the probability of such an extreme event is proportional to $\rho^\beta$, with $\beta=-C\ln{p}/\Gamma>0$. Note that this is actually a lower bound to the small mass tail, which is valid as far as $p\neq0$ and $C\neq0$. Obtaining an analytic expression of the exponent $\beta$ is a theoretically challenging problem that is beyond the scope of the present work. The exponent $\beta$ can be obtained by fitting the left tail of the mass density PDF inside the range where the power law is observed. The fits are presented in dashed lines on the inset of figure \ref{Fig:PDFs_vs_h_b} and the dependence of $\beta$ on the H\"older exponent $h$ is displayed on figure \ref{FIG:exp&var_a} for $\Gamma=10$.

Note that width of the PDF's on figure \ref{Fig:PDFs_vs_h} clearly depends on the value of the H\"older exponent $h$ as it is apparent on figure \ref{FIG:exp&var}.b that represents the variance of the density $\langle \delta\rho^2\rangle=\langle (\rho -\langle\rho\rangle)^2\rangle$. Observe that both the exponent $\beta$ and the variance $\delta \rho$, do not have a monotonic behavior as a function of $h$. This can be qualitatively understood in the following way. When decreasing the value of $h$, the number of zeros of $\sigma(x,t)$ increases. However their spatial distribution also depends on the H\"older exponent $h$. On the one hand, for $h>1/2$ the space increments of $\sigma$ are positively correlated. This implies that there is no accumulation of zeros. For this configuration, the mass is transported from the non-vanishing zones of $\sigma(x,t)$ to the nearest zeros. Therefore decreasing $h$, more and more zeros are present, creating large masses and void zones and thus increasing fluctuations.  On the other hand, for $h<1/2$ the covariance of increments is negative and there are some finite-size regions with a large number of zeros. $\sigma(x,t)$ vanishes many times in such zones, so that the diffusion is very weak and mass is trapped. This makes the transfer inefficient and reduces the probability of having extremely large or low mass concentrations. In simpler words, a large number of zeros increases fluctuations as long as they are not too dense. We expect then a change of behavior near of $h=1/2$,  this is in agreement with figure \ref{FIG:exp&var}.a-b, especially for the largest value of $\Gamma$ where the properties of the system are expected to depend more strongly on the ejection rate distribution of zeros.

\section{Scale invariance of the mass density field\label{Sec:ScaleInv}}

\begin{figure}[h!]
\centering
\subfigure[$h=2$]{\includegraphics[width=.32\columnwidth]{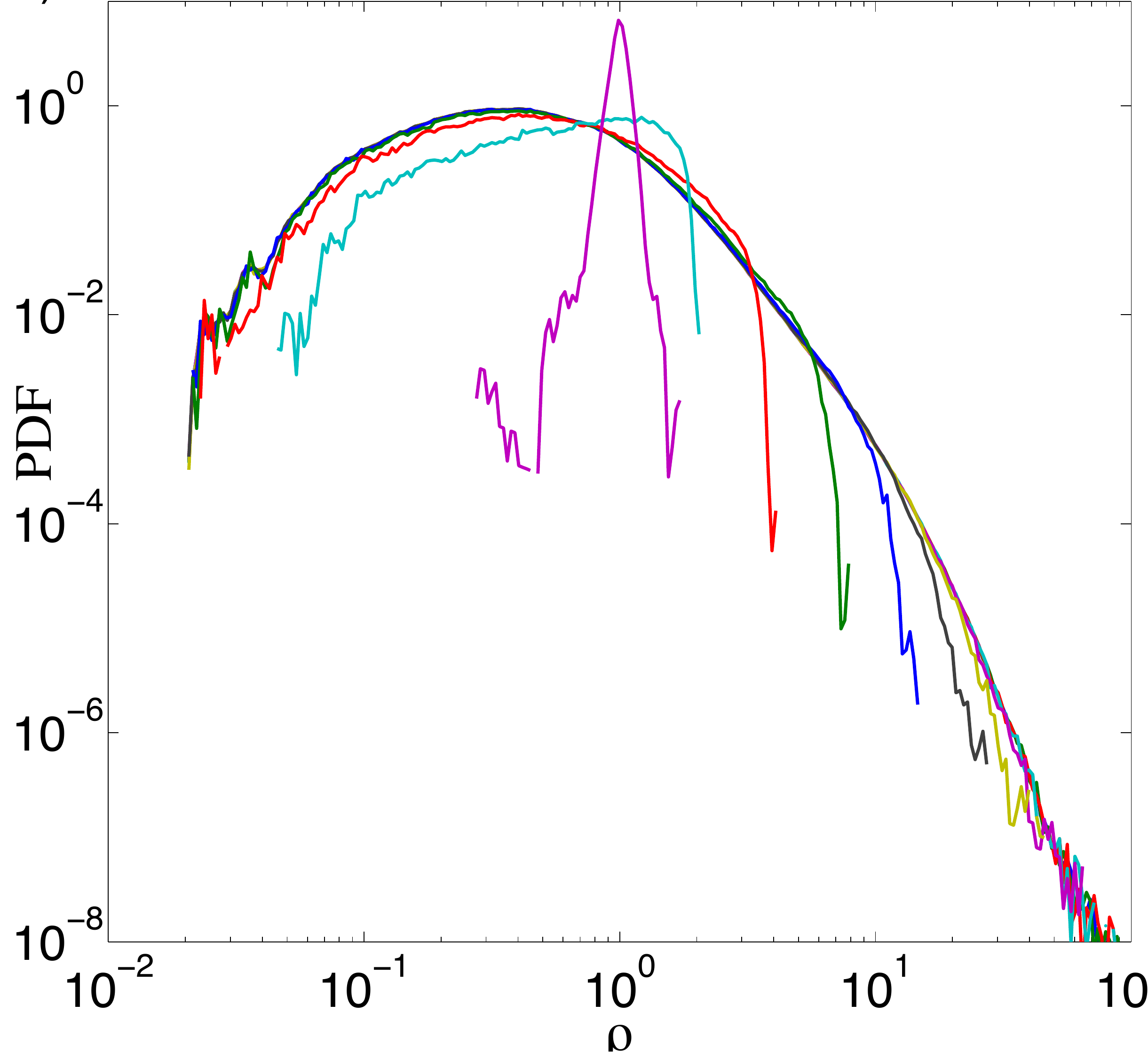}\label{Fig:CGProperties_a}}
\subfigure[$h=0.5$]{\includegraphics[width=.32\columnwidth]{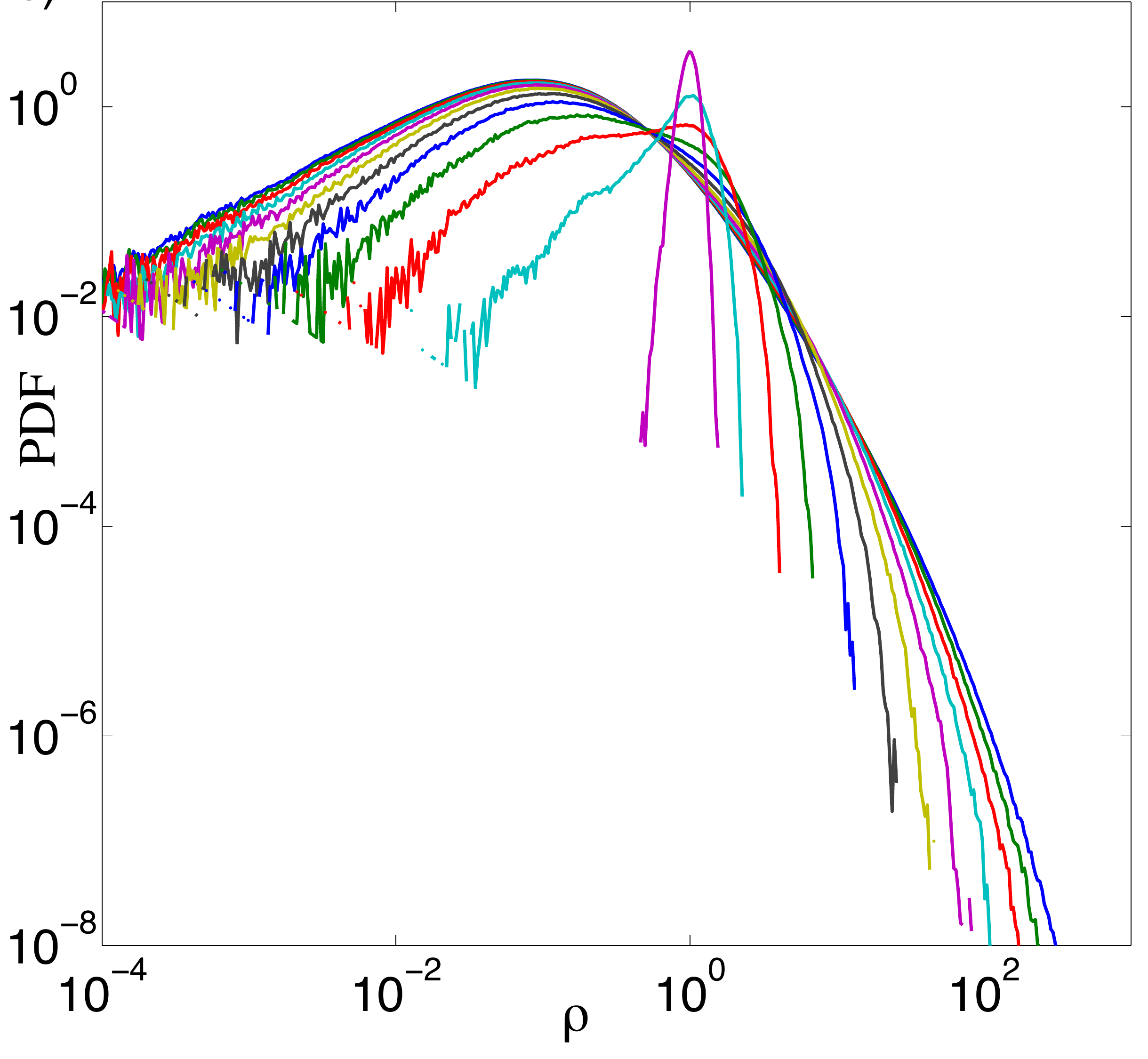}\label{Fig:CGProperties_b}}
\subfigure[$h=0.1$]{\includegraphics[width=.32\columnwidth]{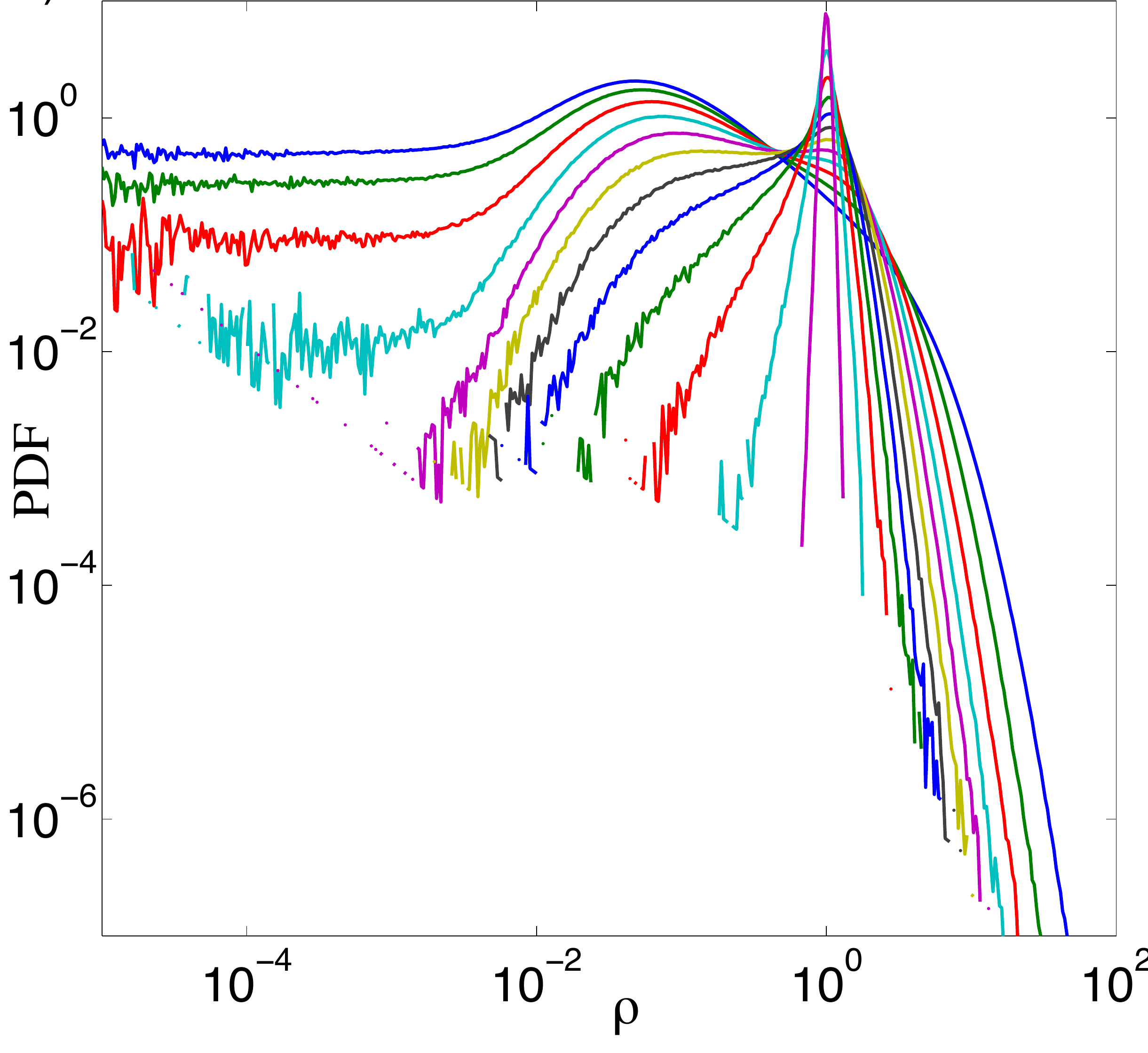}\label{Fig:CGProperties_c}}
 \caption{\label{Fig:CGProperties} PDFs of the density $\rho$ for $\Gamma=1$, different $h$ and various coarse-graining scales. The narrowest curve on each figure corresponds to the largest scale $L/2$ and the curves monotonically increase their width when decreasing the scale down to the smallest scale of the runs, i.e.\ $L/2^{13}$.}
\end{figure}

In this section we study the scaling properties of the density field. For that we introduce the coarse-grained mass density
\begin{equation}
\rho_\ell(x,t)=\frac{1}{\ell}\int_{x-\ell/2}^{x+\ell/2}\rho(y,t)\,\mathrm{d}y. \label{Eq:DefRhoCG}
\end{equation}
Note that by homogeneity we have $\langle \rho\rangle= \langle \rho_\ell\rangle$. The PDF of $\rho_\ell$ is computed using the definition \eqref{Eq:defP} and it is displayed on figure \ref{Fig:CGProperties} for different values of $h$ and $\Gamma=1$. 
The narrowest curves on each figure correspond to the largest (non trivial) scale $L/2$ and the curves monotonically increase their width when decreasing the scale down to the smallest scale of the runs.

For spatially smooth environments, the coarse-grained density of mass becomes invariant when decreasing the scale, as it is apparent on figure \ref{Fig:CGProperties_a} for $h=2$. This indicates that the density field is spatially smooth and that the zeros are isolated. The collapse is faster for small densities. This is due to the accumulation of mass only in some very small clusters and the creation of large void zones. This large voids dominate the coarse-grained density statistics up to their typical size, that is for $\ell<L/8$, as seen from figure \ref{Fig:CGProperties_a}.  On the contrary, the typical size $\ell_\mathrm{clust}$ of mass clusters is rather small. Averaging over scales larger than that this length scale reduces the largest mass fluctuations. Indeed, the coarse-grained density cannot exceed $\ell_\mathrm{clust}/\ell$ and this cutoff decreases with $\ell$, as can be observed on figure \ref{Fig:CGProperties_a} .

\begin{figure}[h!]
\centering
\subfigure[]{\includegraphics[width=.4\columnwidth]{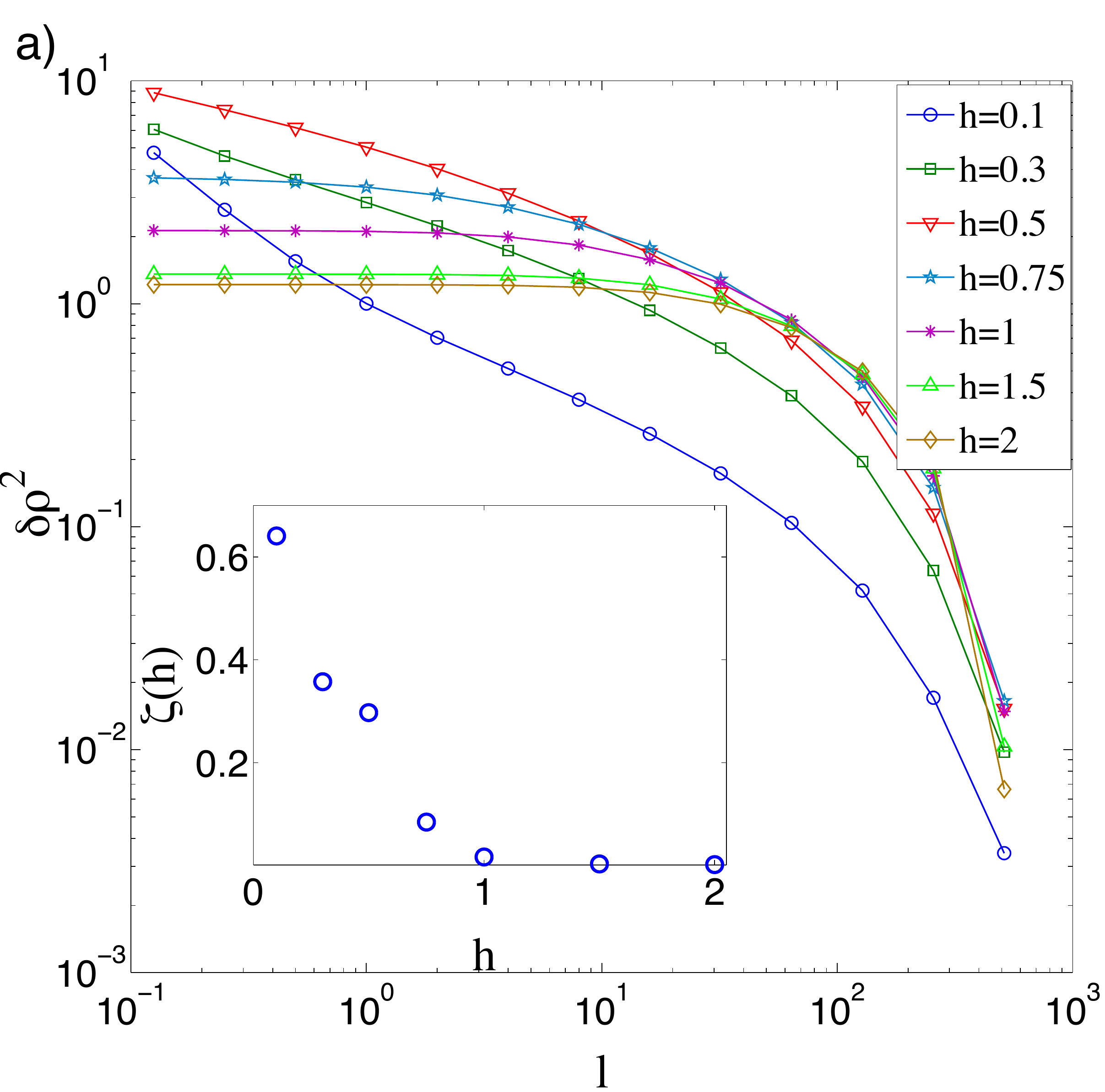}\label{Fig:ScaleInvariance_a}}
\qquad
\subfigure[]{\includegraphics[width=.4\columnwidth]{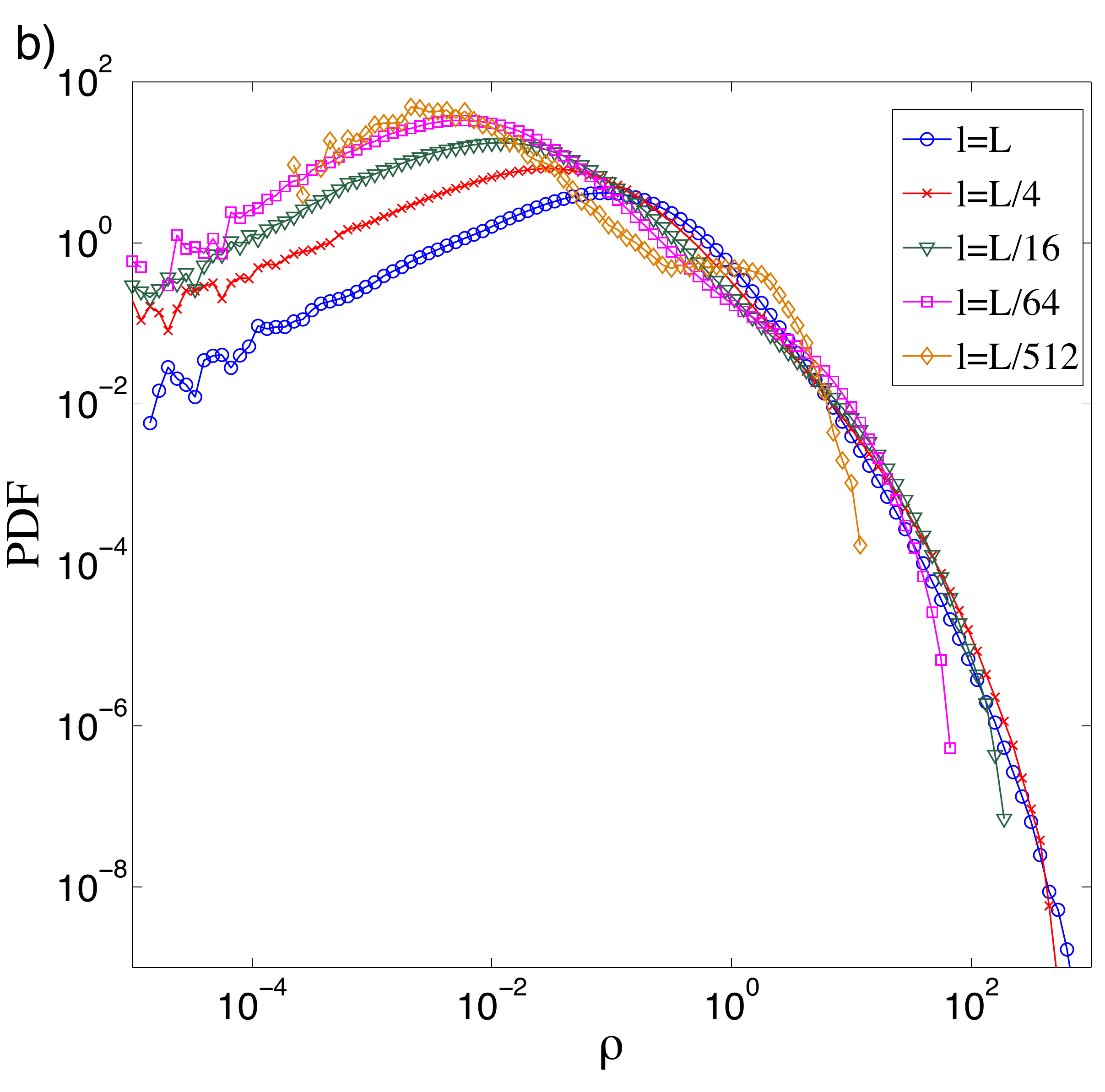}\label{Fig:ScaleInvariance_b}}
\caption{\label{Fig:ScaleInvariance}. (a) Variance  $\langle \delta \rho_\ell^2\rangle=\langle (\rho_\ell-\langle\rho\rangle)^2\rangle$ as a function of $\ell$ for different values of $h$, as labeled. Inset: exponent $\zeta(h;\Gamma)$ of the variance $\langle \delta \rho_\ell^2\rangle \sim \ell^{-\zeta(h;\Gamma)}$ for $\Gamma=1$. b) PDF of the coarse-grained density $\rho_\ell$ for $h=1/2$ and different values of $\Gamma$ and $\ell$ showing a collapse at large masses when the scaling \eqref{Eq:Modelscaled} is used.}
\end{figure}
This argument does not applies for non-smooth environments where the distribution of zeros is highly non trivial. The scale invariance is then broken, as observed on figures \ref{Fig:CGProperties_b} and \ref{Fig:CGProperties_c}. This is also apparent on figure \ref{Fig:ScaleInvariance_a} which represents the variance $\delta\rho_\ell^2=\langle (\rho_\ell-\langle\rho\rangle)^2\rangle$ as a function of $\ell$ for different values of $h$. For $h>1$ and $\ell\ll L$ the variance $\delta\rho_\ell^2$ does not depend on the scale $\ell$ (see the inset of the figure). However, one observes that for $h<1$ and $\ell\ll L$ the variance presents a power-law scaling $\delta\rho_\ell^2 \sim \ell^{-\zeta}$ that suggests a self-similar behavior of the density. The exponent $\zeta(h;\Gamma)$ is displayed on the inset of figure \ref{Fig:CGProperties_c}. 

The scaling invariance of the density field can be understood in the neighborhood of the zeros of $\sigma$. Let us assume that the ejection rate vanishes at $x=x_0$. Then, in the neighborhood of $x_0$, we have $\sigma(x,t)^2=|\sigma(x_0+x,t)|^2\sim x^{2h}$. Supposing that the density behaves as a power-law in the vicinity of $x_0$ and rescaling space as $\tilde{x}=\lambda x$, one can easily see that the  rescaled density $\tilde{\rho}=\rho(\tilde{x})$ is a solution of
\begin{equation}
\pder{\tilde{\rho}(\tilde{x},t)}{t}=\Gamma\lambda^{2-2h}\frac{\partial^2}{\partial \tilde{x}^2}\left[\sigma^2(\tilde{x},t)\tilde{\rho}(\tilde{x},t)\right]\label{Eq:Modelscaled}.
\end{equation}
Hence, we expect that if the coarse-grained density $\rho_\ell$ presents a self-similar property, then the PDF of $\rho_\ell$ obtained with a given value of $\Gamma$ will coincide at large values with that of $\rho_{\lambda\ell}$ corresponding to an ejection rate of amplitude $\Gamma\lambda^{2-2h}$. This scale invariance is apparent on figure \ref{Fig:ScaleInvariance}.b where different PDF's of $\rho_\ell$, with different values of $\Gamma$ and $\ell$ such that $\Gamma\ell^{2h-2}=\mathrm{const}$  are confronted for $h=1/2$ and collapse at large densities.

\section{Conclusions\label{Sec:Concl}}

To summarize, we have introduced a diffusion model based on a simple discrete mass ejection process where the ejection rates are random variables with temporal and spatial correlations. The model can be interpreted in terms of random walks in a time-dependent random environment. We have considered space-periodic environments where the temporal dependence of each Fourier mode is given by independent Ornstein--Uhlenbeck processes and both the amplitudes of the modes and their correlation times present some prescribed spatial scaling properties. This allowed us to consider smooth and non-smooth environments with fast and slow temporal dependence. No other assumptions were made on the environment and we expect such environments to display generic properties and to be representative of sufficiently general situations.

The model was studied analytically and numerically. The corresponding particle dynamics is given by an It\^o differential equation with a multiplicative noise and no drift. We observed that random trajectories diffuse at large times. Also, the probability distribution of their displacements tends to a Gaussian at large times and the deviations to this asymptotic behavior decrease as $t^{-1}$. To take into account the fluctuations of mass due to the randomness of the environment we introduced and studied the probability distribution of the density of mass. We obtained some analytical results on the density of mass in the case of a stationary environment and showed that it displays a power-law tail at large masses with exponent $-2$. In the general case, we observed a competition between a trapping effect due to vanishing ejection rates of the random environment and the mixing due to its temporal dependence which leads to large fluctuations of the density of mass. These fluctuations were studied for both, smooth and non-smooth random environment. In the smooth case, we showed that the PDF has an intermediate power-law behavior, like in the stationary case but with an exponent $-3/2$, followed by an exponential cutoff. Finally, we studied the spatial scaling properties of the mass distribution by introducing a coarse-grained density field. For smooth random environments the coarse-grained density was found to be scale invariant. We showed that at large masses, it possesses some scaling properties that depend on the coarse-graining size, the ejection rate typical amplitude, and the H\"older exponent of the environment.

The overall dynamics of the model that was introduced in this paper contains several space and time scales. Mass rapidly accumulates near the regions with a vanishing ejection rate and slowly moves following the diffusion of the zeros. Depending on the properties of the environment, there is also a clear separation of length scales: small mass fluctuations at small scales and large ones at the scale of the distance of two vanishing ejection rates. This scale separation strongly suggests trying to determine by standard homogenization techniques, a large-scale effective diffusion tensor and eventually an effective transport term. This matter is kept for future work.

Extending our approach to dimensions higher than one is another possible direction for future. The ejection rate will then vanish on complicated sets that, depending on its regularity, might display fractal properties. Varying the H\"older exponent of the random environment, its amplitude and the observation scales could then lead to a rather rich collection of different regimes. For instance, a particular attention would be needed to understand whether or not a power-law is still present in the mass density probability distribution.

To conclude, let us stress again that the zeros of the environment play a crucial role on both the diffusive properties and the mass density statistics. They are responsible for mass accumulation but, at the same time, constitute barriers to transport. This duality implies that the fine statistical details of such systems crucially depend on the features of such zeros and in particular on the local structure of the ejection rate in their vicinity. In this work, we have decided to focus on ejection rates that can be written as the square of a generic Gaussian random function. However this choice cannot be always relevant and, for instance, it might be sometimes more appropriate to write the ejection rate rather as the exponential of a random function. This would change drastically most of the results reported in this paper.

\ack
We warmly acknowledge interesting and stimulating discussions with Beno\^{\i}t Merlet. The research leading to these results has received funding from the European Research Council under the European Community's Seventh Framework Program (FP7/2007-2013 Grant Agreement no.~240579).

\section*{References}
\bibliographystyle{iopart-num}

\begin{thebibliography}{10}
\expandafter\ifx\csname url\endcsname\relax
  \def\url#1{{\tt #1}}\fi
\expandafter\ifx\csname urlprefix\endcsname\relax\def\urlprefix{URL }\fi
\providecommand{\eprint}[2][]{\url{#2}}

\bibitem{Sherrington1975}
Sherrington D and Kirkpatrick S 1975 Wave propagation in random media {\em Phys.\ Rev.\ Lett.\/} {\bf 35}
  1792--1796

\bibitem{HalpinHealy1995215}
Halpin-Healy T and Zhang Y~C 1995 Kinetic roughening phenomena, stochastic growth, directed polymers and all that {\em Phys.\ Rep.\/} {\bf 254} 215--414

\bibitem{ImbrieandSpencer_1988}
Imbrie J~Z and Spencer T 1988 Diffusion of directed polymers in a random environment {\em J.\ Stat.\ Phys.\/} {\bf 52} 609--626

\bibitem{Howe_1971}
Howe M~S 1971 Wave propagation in random media {\em J.\ Fluid Mech.\/} {\bf 45} 769--783

\bibitem{Hughes_RWRE}
Hughes B~D 1996 {\em {Random Walks and Random Environments: Volume 2: Random
  Environments}\/} (Oxford University Press)

\bibitem{solomon1975random}
Solomon F 1975 Random walks in a random environment {\em Annals Prob.\/} {\bf 3} 1--31

\bibitem{sinai1982limit}
Sinai Y 1982 The limit behavior of a one-dimensional random walk in a random environment {\em Teor. Veroyatnost. i Primenen\/} {\bf 27} 247--258

\bibitem{derrida1982classical}
Derrida B and Pomeau Y 1982 Classical diffusion on a random chain {\em Phys.\ Rev.\ Lett.\/} {\bf 48} 627--630

\bibitem{bricmont1991random}
Bricmont J and Kupiainen A 1991 Random walks in asymmetric random environments {\em Comm.\ Math.\ Phys.\/}
  {\bf 142} 345--420

\bibitem{sznitman2002topics}
Sznitman A 2002 Topics in random walks in random environment {\em Notes of
  a Course at the School and Conference on Probability Theory\/} pp 203--266

\bibitem{zeitouni2004part}
Zeitouni O 2004 Random walks in random environment {\em Lectures on probability theory and statistics\/}, Lecture Notes in Mathematics {\bf 1837} (Springer) pp. 191--312

\bibitem{Varadhan2004}
Varadhan S 2004 Random walks in a random environment {\em Proceedings Mathematical Sciences\/} {\bf 114} 309--318

\bibitem{boldrighini2004random}
Boldrighini C, Minlos R and Pellegrinotti A 2004 Random walks in quenched iid space-time random environment are always as diffusive {\em Prob.\ Theory Relat.\ Fields\/} {\bf 129} 133--156

\bibitem{berad2004almost}
B{\'e}rard J 2004 The almost sure central limit theorem for one-dimensional nearest-neighbour random walks in a space-time random environment {\em J. Appl.\ Prob.\/} {\bf 41} 83--92

\bibitem{Frisch:TurbuBook}
Frisch U 1995 {\em {Turbulence : the legacy of A.N. Kolmogorov}\/} (Cambridge University Press)

\bibitem{goudon2004homogenization}
Goudon T and Poupaud F 2004 Homogenization of transport equations: weak mean field approximation {\em SIAM J. Math.\ Analysis\/} {\bf
  36}(3) 856--881

\bibitem{Bec:NJP2007}
Bec J and Ch\'etrite R 2007 Toward a phenomenological approach to the clustering of heavy particles in turbulent flows {\em New J.\ Phys.\/} {\bf 9} 77

\bibitem{OrnsteinUhlenbeck}
Uhlenbeck G~E and Ornstein L~S 1930 On the Theory of the Brownian motion {\em Phys. Rev.\/} {\bf 36} 823--841

\bibitem{mandelbrot:422}
Mandelbrot B~B and Ness J~W~V 1968 Fractional Brownian motions, fractional noises and applications {\em SIAM Rev.\/} {\bf 10} 422--437

\bibitem{feller_1971}
Feller W 1971 {\em An introduction to Probability Theory\/} Vol 1 and 2 (Wiley)

\end{thebibliography}
\providecommand{\newblock}{}

\end{document}